\documentclass[a4paper,11pt]{article}
\pdfoutput=1 

\usepackage{jheppub} 
\usepackage[T1]{fontenc} 
\usepackage{physics}
\usepackage{url}
\usepackage{xcolor}
\usepackage{graphicx}
\usepackage{dcolumn}
\usepackage{bm}

\title{\boldmath Gravitational waves from bubble collisions in FLRW spacetime}

\author[a,b,1]{Haowen Zhong,\note{Corresponding author}}
\author[a]{Biping Gong,}
\author[a,1]{Taotao Qiu}

\affiliation[a]{School of Physics, Huazhong University of Science and Technology\\ Wuhan, 430074, China}
\affiliation[b]{School of Physics and Astronomy, University of Minnesota\\ Minneapolis, MN 55455, USA}
\emailAdd{zhong461@umn.edu}
\emailAdd{bpgong@hust.edu.cn}
\emailAdd{qiutt@hust.edu.cn}

\abstract{Stochastic gravitational wave background (SGWB) is a promising tool to probe the very early universe where the standard model of particle physics and cosmology are connected closely. As a possible component of SGWB, gravitational waves (GW) from bubble collisions during the first order cosmological phase transitions deserve comprehensive analyses. In 2017, Ryusuke Jinno and Masahiro Takimoto proposed an elegant analysis approach to derive the analytical expressions of energy spectra of GW from bubble collisions in Minkowski spacetime avoiding large-scale numerical simulations for the first time\cite{Jinno_2017}. However, they neglect the expansion of the universe and regard the duration of phase transitions as infinity in their derivation which could deviate their estimations from true values. For these two reasons, we give a new expression of GW spectra by adopting their method, switching spacetime background to FLRW spacetime, and considering a finite duration of phase transitions. By denoting $\sigma$ as the fraction of the speed of phase transitions to the expansion speed of the universe, we find when $\sigma$ is around $\mathcal{O}(10)$, the maxima of estimated GW energy spectra drop by around 1 order of magnitude than the results given by their previous work. Even when $\sigma=100$, the maximum of GW energy spectrum is only $65\%$ of their previous estimation. Such a significant decrease may bring about new challenges for the detectability of GW from bubble collisions. Luckily, by comparing new spectra with PLI (\textit{power-law integrated}) sensitivity curves of GW detectors, we find that the detection prospect for GW from bubble collisions is still promising for DECIGO, BBO, LISA, and TianQin in the foreseeable future.}

\begin{document}
\maketitle
\flushbottom

\section{\label{sec:1}Introduction}
Since the first gravitational wave signal has been detected by LIGO on September 14$^{\text{th}}, 2015 $\cite{GW150914}, the study of GW is becoming more and more popular and important. As a brand new detection tool, GW plays a vital role to study  astrophysics and cosmology in this golden era of multi-messenger astronomy. Although the GW events we have detected are all generated from the inspiral-merger-ringdown processes of binary systems, there exists abundant different GW sources especially in the very early universe which inspire our curiosity to a great extent, see Ref.\cite{Caprini_2018} and Ref.\cite{Maggiore_2018} for a comprehensive understanding of the whole picture. The superposition of tremendous amount of GW signals from the very early universe constitutes a stochastic gravitational wave background, where the GW from the first order cosmological phase transitions is a possible component and has been attracting us for a long time because of its relationship to the physics beyond standard model and also the formation of Primordial black holes(PBH), \textit{e.g.} \cite{Espinosa_2008,Ashoorioon_2009,Amjad_2015,Amjad_2021, Das_2010,Huang_2016,Jinno_2017_2,Maxim_1998,Maxim_1999,Maxim_2008}. 

Noticing that GW is a promising tool to probe the very early universe, which in turn is an ideal place to study particle physics, studying GW is of great significance from many different perspectives. As we know, the highest energy scale of LHC (Large Hardron Collider) \cite{LHC} is around 10 TeV which is much less than the energy scale in the very early universe. So there has been no direct means to study particle physics processes happened in that period of time in earth-based laboratories in the foreseeable future. Besides, according to the thermal history of the universe, the oldest photons we can receive are CMB (Cosmic Micro Background) photons which were free from the Thomson scattering and propagated in the universe without restriction when the redshift was around 1100. However, the very early universe we want to study corresponds to the time when the redshift was much larger than 1100. From this point of view, we cannot use electromagnetic waves to probe the universe at that era, because the universe hasn't been transparent for photons at that time. Nevertheless, GW provides a special probe to it from a definitely different dimension. After the generation of GW, it has nearly no interaction with the contents of the universe\cite{Maggiore_2008}, so GW contains a huge amount of precious information from the very early universe compelling us to explore the cosmology and particle physics by studying the properties of GW generated at that epoch.

The first order cosmological phase transitions proceed by the nucleation, expansion and collision of bubbles, at the same time, amount of GW signals are generated. Although nowadays people pay more attention to GW from sound waves\cite{Mark_2014,Mark_2015,Mark_2017} and regard it as the main source of GW from the first order phase transitions, there still are cases like runaway transitions, where the energy density stored in the scalar field can play a really important role i.e. the study of GW from bubble collisions cannot be neglected. The study on GW from bubble collisions needs to be traced back to 1990s, Kosowsky \textit{et al.} did numerical simulations of bubble collisions in Minkowski spacetime with thin wall approximation and envelope approximation and gave energy spectra of GW \cite{Kosowsky_1992_1,Kosowsky_1992_2,Kosowsky_1993,Kosowsky_1994}. The 
subsequent works by others using numerical simulations can refer to Ref.\cite{Huber_2008, Weir_2018}.  However, considering that large-scale numerical simulations spend lots of computational resources and time, people have been trying to understand the physics of bubble collisions from an analytical view as well. In 2008, Caprini \textit{et al.} proposed an alternative analytical method to study the GW from bubble collisions relying on a different treatment of the nature of stochasticity\cite{Caprini_2008}. In 2017, Caprini's ansatz for correlator function has been adopted by Jinno and Takimoto\cite{Jinno_2017} who developed a brand new analysis method to study GW spectra from bubble collisions in Minkowski spacetime by studying the past cones of the spacetime points $x$ and $y$ which show up in two-point correlator of energy-momentum tensor $\langle T_{ij}(x)T_{ij}(y)\rangle$. Their method not only gives a perspicuous physical picture, but also saves computational sources by avoiding large-scale numerical simulations and averts accompanying numerical errors. However, their work is based on Minkowski spacetime rather than FLRW spacetime which is definitely a better choice to take the expansion of the universe into account. Besides, they neglect the effect brought about by the finite duration of the phase transitions which could make the energy density of GW they estimated larger than the real value.

In this paper, we discuss GW from bubble collisions during the first order cosmological phase transitions in FLRW spacetime during RD(\textit{Radiation Dominated}) era via the analysis method proposed by Jinno and Takimoto in Ref.\cite{Jinno_2017} and also take the effect of finite duration of phase transitions into account. As the analyses in Minkowski spacetime, the core quantity for GW spectrum calculation is still the transverse and traceless part of two-point correlator of energy-momentum tensor $\langle T_{ij}(x)T_{ij}(y)\rangle$. With the help of thin wall approximation and envelope approximation, we successfully obtain the analytical expressions of GW spectra described by two triple integrations which can be integrated numerically. By comparing the GW spectra with PLI (\textit{power-law integrated}) sensitivity curves of several space-borne GW detectors and pulsar timing arrays, we can estimate if GW from bubble collisions in FLRW spacetime during RD era can be detected or not, which has important practical significance for future detection. Although our derivation is limited to the phase transitions happened in RD era, the analytical expressions of GW spectra we obtain is model-independent, and the derivation in other eras can be extended easily by specifying the relationship between Hubble parameters, conformal time and scale factor. 

The organization of this paper is as follows. In Sec.\ref{sec:2}, a brief retrospection and summary of Jinno and Takimoto's work is given. In Sec.\ref{sec:3}, we introduce the basic ingredients that the subsequent analyses need, including our assumptions and derivations of GW spectra. In Sec.\ref{sec:4}, we obtain the analytical expressions of $\Delta^{\text{F}(s)}$ and $\Delta^{\text{F}(d)}$ which can both be written as a triple integration, where $\Delta^{\text{F}(s)}$ and $\Delta^{\text{F}(d)}$ are defined in Sec.\ref{sec:2}. At the end of this section, we also prove our integration expression of $\Delta^{\text{F}}$ can return to Jinno and Takimoto's expression analytically when $\sigma\to\infty$. In Sec.\ref{sec:5}, we introduce our method to determine the “effective duration” of phase transitions and then show the numerical estimations of $\Delta^{\text{F}(s)}$ and $\Delta^{\text{F}(d)}$ and a comparison between our results and the results obtained in Minkowski spacetime. To testify the detectability of GW, we compare GW spectra with PLI sensitivity curves of several GW detectors in the end of Sec.\ref{sec:5}. A thorough conclusion of this work and some extra discussion are given in Sec.\ref{sec:6}.
\section{\label{sec:2}A Brief Retrospection of Jinno and Takimotos' Work}

In this section, we give a brief retrospection and summary of their work. Because our analyses and calculations are all based on their model, we summarize the assumptions and approximations adopted by them at here and directly list the result they've obtained. The calculation details won't be shown in this section, so for those readers who are interested in those details can read Ref.\cite{Jinno_2017} to get a more in-depth understanding.

Their calculations are based on Minkowski spacetime, including tensor perturbation the metric can be written as :
\begin{equation}
    \text{d}s^2=-\text{d}t^2+(\delta_{ij}+2h_{ij})\text{d}x^i\text{d}x^j
\end{equation}
where $h_{ij}$ are all transverse and traceless. The speed of light $c$ has been set to 1, we'll adopt this convention as well in our own derivation. As for the convention of indices, the Greek indices run over $0,1,2,3$ and the Latin indices run over $1,2,3$ throughout our paper.
\subsection{Assumptions and approximations}
They adopt thin wall approximation and envelope approximation in their work. The so-called thin wall approximation means that we assume all energy of the bubble is located in the bubble wall with an infinitesimal width $\ell_B$. With the help of this approximation, they write the energy momentum tensor $T^B$ of the uncollided wall of a single bubble nucleated at $x_N\equiv(t_N,\bm{x}_N)$ as:
\begin{equation}
    T_{ij}(x)=\rho(x)\widehat{(x-x_N)}_i\widehat{(x-x_N)}_j
\end{equation}
where
\begin{equation}
\rho(x)=\left\{
\begin{aligned}
&\frac{\frac{4\pi }{3}r_B(t)^3\kappa\rho_0}{4\pi r_B(t)^2\ell_B} \quad   r_B(t)<\abs{\bm{x}-\bm{x}_N}<r_B'(t)\\
&\hspace{1cm}0\hspace{2.3cm} \text{Other regions}
\end{aligned}
\right.
\end{equation}
and 
\begin{equation}
    r_B(t)=v(t-t_N)\quad r_B'(t)=r_B(t)+\ell_B
\end{equation}
At here, $x\equiv(t,\bm{x})$ denotes the spacetime point, $\hat{\bullet}$ indicates the unit vector in the same direction of $\Vec{\bullet}$, $v$ is the speed of the bubble wall, $\rho_0$ represents the energy density released by the phase transitions, and $\kappa$ denotes the fraction of the energy localized in the bubble wall to the released energy during the phase transition. 

\begin{figure}
    \centering
    \includegraphics[width=0.45\textwidth]{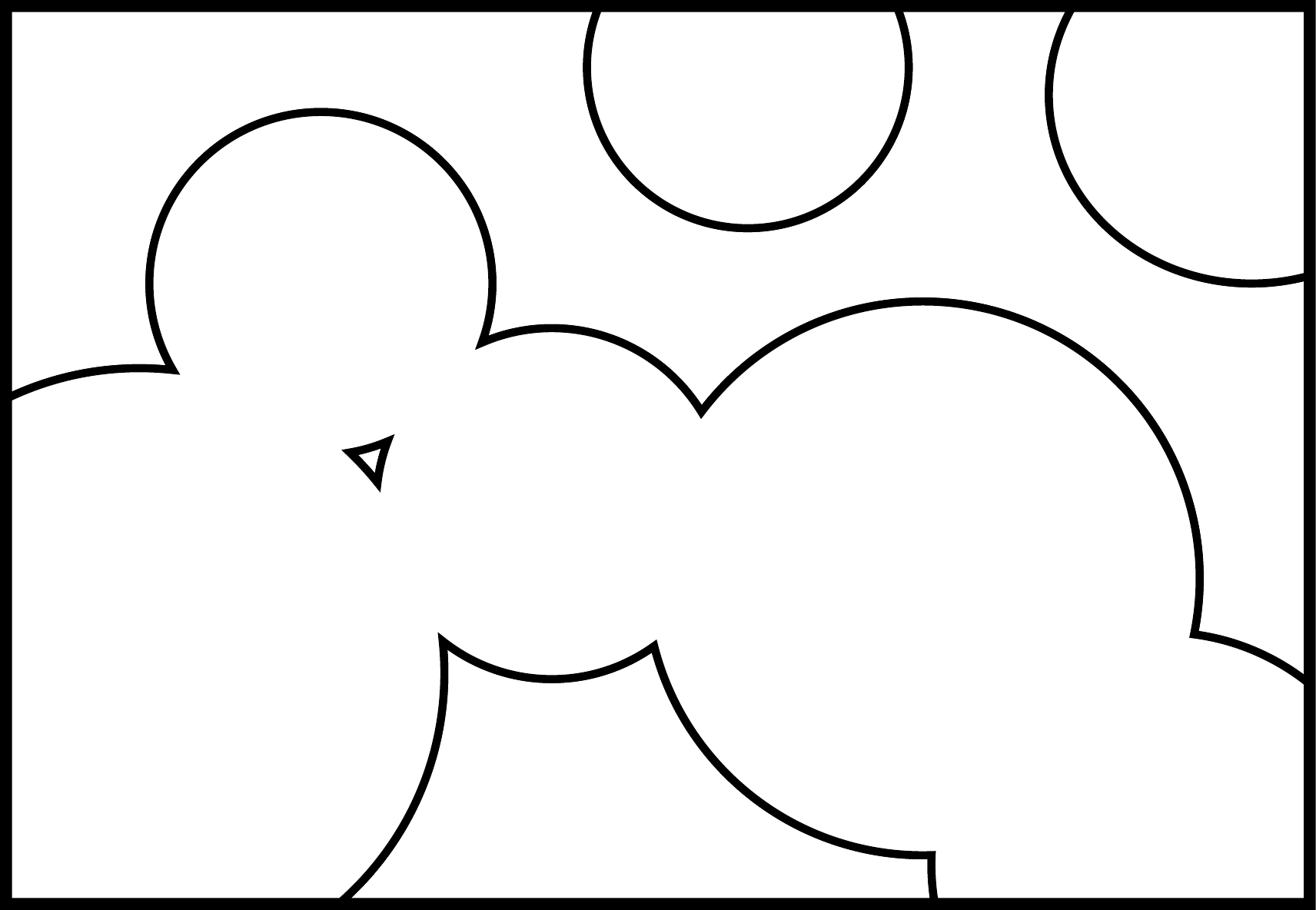}
    \caption{2D explanation of envelope approximation, the black lines represent bubble walls. Only uncollided parts of bubbles are regarded as GW sources, and the energy-momentum tensor of the collided parts of bubbles vanish at once.}
    \label{fig:envelope}
\end{figure}

Envelope approximation means that as long as two or more bubbles collide, the colliding parts of bubble walls and corresponding energy-momentum tensor vanish at once. With the guarantee of envelope approximation, every spatial point($\bm{x}$) can be passed by bubble only once. Since once a spatial point enter the inner part of any single bubble, it will transform from the false vacuum state to the true vacuum state and the inverse process is forbidden.
See FIG.\ref{fig:envelope} for a straightforward understanding of this important approximation.

Besides these two approximations, they denote the bubble nucleation rate per unit time per volume as:
\begin{equation}
    \Gamma(t)=\Gamma_\star \text{e}^{\beta(t-t_\star)}
\end{equation}
where $t_\star$ denotes a time point typically around the transition time, $\Gamma_\star$ is the transition rate at $t_\star$, and $\beta$ is a constant.

In order to obtain the GW spectrum, they define equal-time correlator of GW and unequal-time correlator of energy momentum tensor given by Eq.(\ref{eq:correlator}), which we'll also take use of in our own derivation.
\begin{equation}
\begin{aligned}
    &\langle \dot{h}_{ij}(t,\bm{k})\dot{h}_{ij}^*(t,\bm{q})\rangle =(2\pi)^3\delta^{(3)}(\bm{k}-\bm{q})P_{\dot{h}}(t,k)\\
    &\langle \Pi_{ij}(t_x,\bm{k})\Pi_{ij}^*(t_y,\bm{q})\rangle=(2\pi)^3\delta^{(3)}(\bm{k}-\bm{q})\Pi(t_x,t_y,k)
\end{aligned}
\label{eq:correlator}
\end{equation}
Where $\Pi_{ij}$ is the transverse-traceless part of the energy-momentum tensor $T_{ij}$.
\subsection{GW spectrum}
According to the assumptions and approximations given by the previous subsection, they find the core quantity we need to obtain for describing the GW spectrum is the transverse and traceless part of $\langle T_{ij}(x)T_{ij}(y) \rangle$ i.e. $\langle \Pi_{ij}(x)\Pi_{ij}(y)\rangle$. After figuring it out, we can get $\Pi(t_x,t_y,k)$ with the help of Eq.(\ref{eq:correlator}) and Fourier transformation. As long as we have $\Pi(t_x,t_y,k)$, we can calculate the dimensionless GW energy density per logarithmic frequency $\Omega_{\text{GW}}$ by:
\begin{equation}
\begin{aligned}
    \Omega_{\text{GW}}:=\frac{1}{\rho_{\text{tot}}}\frac{\text{d}\rho_{\text{GW}}}{\text{d}\ln k}&=\kappa^2\Bigg(\frac{H_\star}{\beta}\Bigg)^2\Bigg(\frac{\alpha}{1+\alpha}\Bigg)^2\Delta(k/\beta,v)\equiv\kappa^2\Bigg(\frac{\alpha}{1+\alpha}\Bigg)^2\Delta^{\text{M}}(k/\beta,\sigma,v)
    \end{aligned}
    \label{eq:defDeltaJinno}
\end{equation}
\begin{equation}
    \sigma:=\frac{\beta}{H_\star}\quad\quad\Delta^{\text{M}}(k/\beta,\sigma,v)\equiv\frac{\Delta(k/\beta,v)}{\sigma^2}
\end{equation}
The superscript M indicates Minkowski spacetime. The definition of $\Delta(k/\beta,v)$ is given by
\begin{equation}
    \Delta(k/\beta,v)=\frac{3}{4\pi^2}\frac{\beta^2k^3}{\kappa^2\rho_0^2}\int_{t_i}^{t_f}\text{d}t_x\int_{t_i}^{t_f}\text{d}t_y\cos[k(t_x-t_y)]\Pi(t_x,t_y,k)
\label{eq:Delta_jinno}
\end{equation}
Here $\rho_{\text{tot}}=\rho_0+\rho_{\text{rad}}$, $\alpha:=\rho_0/\rho_{\text{rad}}$ represents the fraction of the released energy density from the phase transitions to the energy density of radiation, $H_\star$ denotes the typical value of Hubble parameter around the transition time, $t_i$ and $t_f$ denote the start time and end time of phase transitions. $\sigma$ is defined as the ratio of $\beta$ and $H_\star$, which can characterize the speed of phase transitions to the expansion speed of the universe. For the convenience of latter comparison, we define a new variable $\Delta^{\text{M}}$ at here, we'll see that $\Delta^{\text{M}}$ is the specific variable that we'll use to compare with our results in Sec.\ref{sec:5} rather than $\Delta$ itself. To prevent possible confusion, we'll explicitly write superscript M to distinguish $\Delta$ and $\Delta^{\text{M}}$. 

\begin{figure}
    \centering
    \includegraphics[width=0.7\textwidth]{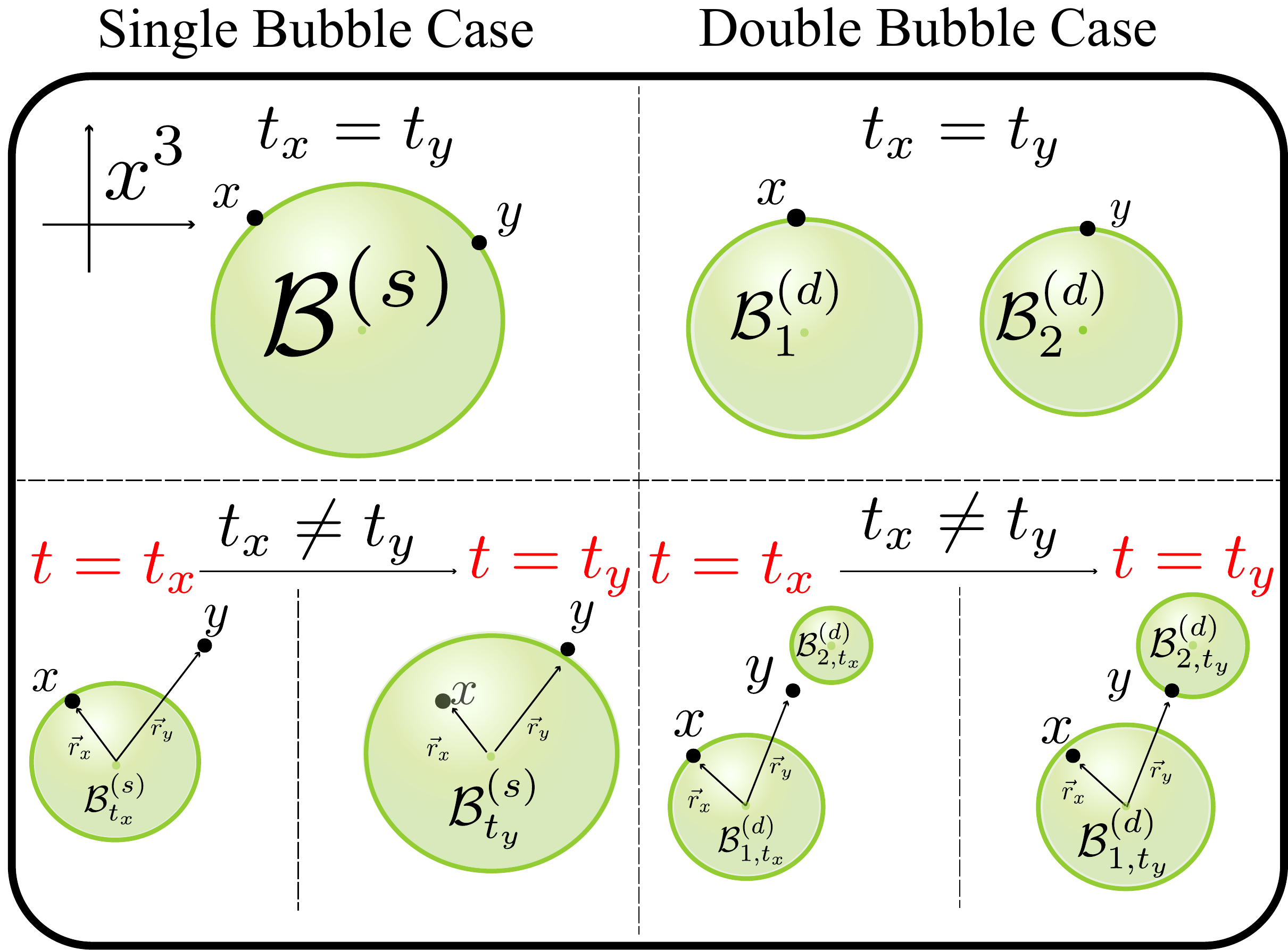}
    \caption{Sketch of single bubble case and double bubble case. $x$ and $y$ are two spacetime points we are studying, bubble(s) is/are denoted by $\mathcal{B}$, the superscripts $(s)$ and $(d)$ distinguish the single bubble case and double bubble case. The subscripts $t_x$ and $t_y$ denote $t=t_x$ or $t=t_y$, subscript $1$ and $2$ denote the first bubble and the second bubble in double bubble case. The left panel corresponds to the single bubble case, and the right panel characterizes the double bubble case. Note that we don't require the spatial points $\bm{x}$ and $\bm{y}$ to stay in the bubble walls at the same time, so they can be in the bubble walls at different time as the lower panel of this figure displays.}
    \label{fig:twocase}
\end{figure}

According to analyses, $\Delta$ can be divided into two parts as $\Delta=\Delta^{(s)}+\Delta^{(d)}$. $\Delta^{(s)}$ and $\Delta^{(d)}$ correspond to the contribution to $\Delta$ from the single bubble case and double bubble case respectively. The so-called single bubble case and double bubble case need to be understood by considering the relative positions of bubble walls and two spatial points $\bm{x}$ and $\bm{y}$. It is easy to realize that only if $T_{ij}(x)\neq0$ and $T_{ij}(y)\neq0$ can this situation contributes to $\langle T_{ij}(x)T_{ij}(y)
\rangle$, which means that the spatial points $\bm{x}$ and $\bm{y}$ must be in the bubble wall(s) at the time points $t_x$ and $t_y$, respectively. So we can find that there are two possible situations. The first one is that $\bm{x}$ and $\bm{y}$ are passed by the same bubble at $t_x$ and $t_y$ respectively, and the second one is that $\bm{x}$ and $\bm{y}$ are passed by two different bubbles at $t_x$ and $t_y$ respectively. FIG.\ref{fig:twocase} displays these two possible situations, the left panel and right panel correspond to the single bubble case and double bubble case, respectively. Because $t_x$ can be equal or not equal to $t_y$, for each case there are two more possibilities. The upper left panel shows the situation where $t_x=t_y$. The spatial points $\bm{x}$ and $\bm{y}$ are passed by the same bubble at $t=t_x=t_y$. The lower left panel shows the situation where the bubble firstly passes the spatial point $\bm{x}$ at $t=t_x$ and lately passes the spatial point $\bm{y}$ at $t=t_y$. The right panel can be understood similarly.

In order to calculate $\Delta^{(s)}$ and $\Delta^{(d)}$, we need to study the past cones of $x$ and $y$ to decide the permitted spacetime points for bubbles to nucleate. For each possible nucleation point, we need to figure out two quantities. The first one is the probability of such a situation, and the second one is the contribution to $\langle T_{ij}(x)T_{ij}(y)\rangle$ from this situation. By summing up all of the possible situations, we can get the final result of $\langle T_{ij}(x)T_{ij}(y)\rangle$. In a word, calculation process can be described by $\displaystyle{\langle \text{TT}\rangle(x,y)=\sum_{\text{situation}}\text{P}(\text{situation})\text{TT}(x,y)\Big|_{\text{situation}}}$. The detailed calculation idea can be found in their Appendix A. 

As a result of their work, Jinno and Takimoto find that the GW spectrum can be described as two double integrations which can be calculated numerically. The shape of GW spectrum they've got is consistent with the result given by numerical simulations. They also give the fitting formula of $\Delta$ as:
\begin{equation}
    \Delta=\frac{\Delta_{\text{peak}}}{c_l\Big(\frac{f}{f_{\text{peak}}}\Big)^{-3}+(1-c_l-c_h)\Big(\frac{f}{f_\text{peak}}\Big)^{-1}+c_h\Big(\frac{1}{f_\text{peak}}\Big)}
    \label{eq:jinnofitting}
\end{equation}
with $\Delta_{\text{peak}}=0.043, f_{\text{peak}}/\beta=1.24/2\pi$ and $(c_l,c_h)=(0.064,0.48)$. We'll use this formula to calculate $\Delta$ and compare it with our results in Sec.\ref{sec:5}.
\section{\label{sec:3}Basic setup}
Because our discussion is based on FLRW spacetime, let's recall the FLRW metric at the very beginning:
\begin{equation}
    \text{d}s^2=a^2(\eta)(-\text{d}\eta^2+\text{d}x^2+\text{d}y^2+\text{d}z^2)
\label{eq:one}
\end{equation}
where $x,y,z$ are comoving coordinates, $a(\eta)$ is scale factor and the argument $\eta$ is conformal time whose definition is given by:
\begin{equation}
    \text{d}\eta:=\frac{\text{d}t}{a(t)}\Longrightarrow\eta:=\int^t\frac{\text{d}t}{a(t)}
\end{equation}
with $t$ being called cosmic time. In the latter parts of this section, we firstly specify the approximations and assumptions we use in this paper, and then we give the GW spectrum at the end of phase transitions and nowadays which we can detect directly.
\subsection{\label{sec:3.1}Assumptions and approximations}
In this work, we adopt thin wall approximation and envelope approximation as Jinno and Takimoto's choice in their work. The concrete definitions of these two approximations have been discussed in Sec.\ref{sec:2}. According to Jinno and Takimoto's setup, we can write the energy-momentum tensor of an uncollided bubble nucleated at $x_N\equiv(\eta_N,\bm{x}_N)$ as:
\begin{equation}
    T_{ij}(x)=a^2(\eta)\Bigg(\rho(x)\widehat{(x-x_N)}_i\widehat{(x-x_N)}_j\Bigg)
    \label{eq:Tij}
\end{equation}
where
\begin{equation}
\rho(x)=\left\{
\begin{aligned}
&\frac{\frac{4\pi}{3}r_B(\eta)^3\kappa\rho_0}{4\pi r_B(\eta)^2a(\eta)\ell_B} \quad r_B(\eta)<\Delta\bm{x}<r_B'(\eta)\\
&\hspace{1cm}0\hspace{2.3cm} \text{Other regions}
\end{aligned}
\right.
\end{equation}
with $\Delta\bm{x}\equiv a(\eta)\abs{\bm{x}-\bm{x}_N}$. To remind, the expressions of $T_{ij}$ given by Eq.(\ref{eq:Tij}) is a modeling rather than a rigorous expression by derivation. The physical meanings of $\rho_0$ and $\kappa$ are both the same as which has been given in the previous section. In our work, we only consider the situation where the speed of bubble walls equals to the speed of the light, i.e. $v=c=1$. So we have:
\begin{equation}
    r_B(\eta)=\eta-\eta_N\quad r_B'(\eta)=r_B(\eta)+\ell_B
\end{equation}
Note that $\ell_B$ is the comoving width of the bubble wall. For the future convenience, we define $\Psi_{ij}(\eta,\bm{x})$ and $\Psi_{ij}(\eta,\bm{k})$ as:
\begin{equation}
       \Psi_{ij}(\eta,\bm{x}):=\frac{T^{\text{TT}}_{ij}(\eta,\bm{x})}{a^2(\eta)}\Longrightarrow \Psi_{ij}(\eta,\bm{k}):=\frac{T^{\text{TT}}_{ij}(\eta,\bm{k})}{a^2(\eta)}
\end{equation}
where superscript TT is the abbreviation of Transverse-Traceless. $\Psi_{ij}(\eta,\bm{k})$ and $T_{ij}^{\text{TT}}(\eta,\bm{k})$ are the Fourier modes of $\Psi_{ij}(\eta,\bm{x})$ and $T_{ij}^{\text{TT}}(\eta,\bm{x})$ with wave vector $\bm{k}$, respectively. In this work, our conventions of Fourier transformation are given by $\displaystyle{\int \text{d}^3xe^{i\bm{k}\cdot\bm{x}}}$ and $\displaystyle{\int\text{d}^3k/(2\pi)^3e^{-i\bm{k}\cdot\bm{x}}}$. We can extract the TT parts of $T_{ij}(\eta,\bm{k})$ by contracting it with Lambda tensor whose definition is given by Eq.(\ref{eq:lambda}).

The transition ratio per unit conformal time per comoving volume we use in our paper is given by\cite{Caprini_2008}:
\begin{equation}
    \Gamma(\eta)=\Gamma_\star e^{\widetilde{\beta}(\eta-\eta_\star)}
    \label{eq:nucleate}
\end{equation}

where $\eta_\star$ denotes the conformal time corresponding to the start of phase transitions, and $\Gamma_\star$ is the value of $\Gamma(\eta)$ at this specific time. Note that the definition of $\widetilde{\beta}$ is given by $\widetilde{\beta}:= a(\eta)\beta\Big|_{\eta=\eta_\star}=$const. where $\beta$ is more common to be used in the literature \textit{e.g.} Ref.\cite{Kosowsky_1993}. At here, it is necessary to explain that Eq.(\ref{eq:nucleate}) only holds when $\eta\sim\eta_\star$, so in our work, we consider a finite length of phase transitions to make sure this relationship can be satisfied. Actually, later we'll see the choice of FLRW spacetime itself gives us a relatively natural way to determine the duration of phase transitions which will directly influence the final GW spectrum. In most literature, people estimate the duration of the phase transitions by $\beta^{-1}$, and in Jinno and Takimoto's work they set the phase transitions start at $t_i=-\infty$ and end at $t_f=+\infty$ which itself means that they regard the duration of phase transitions as infinite long in their actual derivation and numerical calculation. Although they give an argument to this point that this treatment doesn't change their final results, in our view, a finite duration might be a better choice and could make the results more reliable. The exact method to determine the length of phase transitions will be discussed in the latter section.

Without loss of generality, we will set $\widetilde{\beta}\equiv1$ in the rest of this paper which is just a choice of the unit like setting the speed of light $c\equiv1$. Besides, it's also worth mentioning that although $\widetilde{\beta}$ is a constant given a certain value of $\beta$ and $a$ i.e. given a specific physical scenario, we are not allowed to compare any dimensional physical quantity by comparing the number of them in different scenarios. Because the physical meaning of “1” in each 
particular scenario is different. However, by the virtue of dimensionless property of the  quantity $h_0^2\Omega_{\text{GW}}$, we can compare it safely in different physical scenarios.

\subsection{GW power spectrum derivation}
In this part we begin to derive the analytical expressions of GW spectrum using the model proposed by Ref.\cite{Jinno_2017}.

Let's consider the equation of motion of the metric perturbation satisfy at first. As we have stated already, our spacetime background is FLRW spacetime. Including tensor perturbations, we have the spacetime metric as:
\begin{equation}
    \text{d}s^2=a^2(\eta)[-\text{d}\eta^2+(\delta_{ij}+h_{ij})\text{d}x^i\text{d}x^j]
\end{equation}
Here $h_{ij}$ are all traceless and transverse. By calculating the Christoffel symbol, Riemman tensor and Ricci tensor, we have the linearized Einstein's equation to the first order in $h_{ij}$, over the FLRW spacetime as \cite{Caprini_2018}:
\begin{equation}
    \ddot{h}_{ij}(\bm{x},t)+3H\dot{h}_{ij}(\bm{x},t)-\frac{\nabla^2}{a^2}h_{ij}(\bm{x},t)=16\pi G\Psi_{ij}(\bm{x},t)
\label{eq:eom}
\end{equation}
Where $H$ is Hubble parameter at $t$, $\nabla^2\equiv\partial^i\partial_i$ indicates the Laplacian associated with comoving coordinates $x^i$ and $\dot{\bullet}$ denotes the derivative of $\bullet$ respective to cosmic time $t$. Define $H_{ij}(\eta,\bm{x})=a(\eta)h_{ij}(\eta,\bm{x})$ and Eq.(\ref{eq:eom}) in Fourier space becomes
\begin{equation}
    H_{ij}^{\prime\prime}(\bm{k},\eta)+\Bigg(k^2-\frac{a^{\prime\prime}}{a}\Bigg)H_{ij}(\bm{k},\eta)=16\pi Ga^3\Psi_{ij}(\bm{k},\eta)
\end{equation}
where $\bullet^{\prime}$ denotes the derivative of $\bullet$ with respect to conformal time $\eta$. For the convenience of subsequent derivation, we define $\zeta\equiv k\eta$ at here.  Since we assume that the phase transitions happen in RD era, we can use the relationship of scale factor and conformal time $a(\eta)=\chi\eta$ to simplify the equation of motion and obtain Eq.(\ref{eq:eom_sim}) where $\chi$ is a constant.
\begin{equation}
    \frac{\text{d}^2H_{ij}(\bm{k},\zeta)}{\text{d}\zeta^2}+H_{ij}(\bm{k},\zeta)=\frac{16\pi G\chi^3}{k^5}\zeta^3\Psi_{ij}(\bm{k},\zeta)
\label{eq:eom_sim}
\end{equation}

As we have stated before, $\Psi_{ij}(\eta,\bm{x})$ is related to the TT part of energy-momentum tensor $T_{ij}$, and we can use Lambda tensor to extract the TT part of $T_{ij}$. The definition of Lambda tensor is given by:
\begin{equation}
\left\{
\begin{aligned}
&\Lambda_{ij,kl}(\widehat{\bm{k}}):=P_{ik}(\widehat{\bm{k}})P_{jl}(\widehat{\bm{k}})-\frac{1}{2}P_{ij}(\widehat{\bm{k}})P_{kl}(\widehat{\bm{k}})\\
&P_{ij}(\widehat{\bm{k}}):=\delta_{ij}-\widehat{\bm{k}}_i\widehat{\bm{k}}_j
\end{aligned}
\right.
\label{eq:lambda}
\end{equation}
It's easy to check that the Lambda tensor has a basic property which is of great use:
\begin{equation}
\Lambda_{ij,kl}(\widehat{\bm{k}})\Lambda_{ij,mn}(\widehat{\bm{k}})=\Lambda_{kl,mn}(\widehat{\bm{k}})=\Lambda_{mn,kl}(\widehat{\bm{k}})
\end{equation}
So we can rewrite $\Psi_{ij}$ in terms of $T_{ij}$ with the help of Lambda tensor:
\begin{equation}
    \Psi_{ij}(\eta,\bm{k})=\frac{\Lambda_{ij,kl}(\bm{k})T_{kl}(\eta,\bm{k})}{a^2(\eta)}
\end{equation}

Assuming that the phase transitions begin at $\zeta_i=k\eta_i$ and end at $\zeta_f=k\eta_f$. $\Psi_{ij}=0$ before the phase transitions. Let's consider a time point $\zeta_i<\zeta<\zeta_f$ during the phase transitions, we can get the solution of $H_{ij}$ by Green's function method:
\begin{equation}
    H_{ij}(\bm{k},\zeta<\zeta_f)=\frac{16\pi G\chi^3}{k^5}\int_{\zeta_i}^{\zeta}\text{d}\psi\ \psi^3\sin(\zeta-\psi)\Psi_{ij}(\bm{k},\psi)
    \label{eq:solution_Green}
\end{equation}
When phase transitions finish, $\Psi_{ij}=0$ (no bubbles anymore,  all spacetime points stay in the true vacuum state), so we have
\begin{equation}
    H_{ij}(\bm{k},\zeta>\zeta_f)=A_{ij}(\bm{k})\cos \zeta+B_{ij}(\bm{k})\sin \zeta
\label{eq:solution}
\end{equation}
We require $H_{ij}$ to satisfy the connection requirements at $\zeta=\zeta_f$, i.e. $H_{ij}|_{\zeta_f}$ and $H^{\prime}_{ij}|_{\zeta_f}$ given by Eq.(\ref{eq:solution_Green}) and Eq.(\ref{eq:solution}) should be equal to each other. Using these requirements, we can arrive at
\begin{equation}
\left\{
\begin{aligned}
&A_{ij}(\bm{k})=\frac{16\pi G\chi^3}{k^5}\int_{\zeta_{i}}^{\zeta_{f}}\text{d}\psi\ \psi^3\sin(-\psi)\Psi_{ij}(\bm{k},\psi)\\
&B_{ij}(\bm{k})=\frac{16\pi G\chi^3}{k^5}\int_{\zeta_{i}}^{\zeta_{f}}
\text{d}\psi\ \psi^3\cos(\psi)\Psi_{ij}(\bm{k},\psi)
\end{aligned}
\right.
\label{eq:ab}
\end{equation}
Insert Eq.(\ref{eq:ab}) into Eq.(\ref{eq:solution}), we can get the final expression of $H_{ij}$. Note that this solution only holds during the RD era, so our results cannot be extended to the phase transitions that happen in MD (\textit{Matter Dominated}) or $\Lambda$D (\textit{$\Lambda$ Dominated}) eras.

Now let's discuss the power spectrum of GW. First of all, we define the equal-time two-point correlator of GW by
\begin{equation}
\langle h'_{ij}(\eta,\bm{k})(h'_{ij})^*(\eta,\bm{q})\rangle:=(2\pi)^3\delta^{(3)}(\bm{k}-\bm{q})P_{h'}(\eta,k)
\label{eq:eqcor}
\end{equation}
Where $\langle...\rangle$ denotes ensemble average and we will use spatial and time average to substitute ensemble average in the actual calculation. $(2\pi)^3$ is a coefficient putting at here out of convenience for future derivation. $\delta^{(3)}(\bm{k}-\bm{q})$ shows up out of the homogeneity of the system. Interested readers could refer to Chapter 7 of Ref.\cite{Maggiore_2008} to know more details and more properties of stochastic gravitational wave background.

We define the unequal-time correlator of energy-momentum tensor by\footnote{For simplicity, we call $T_{ij}$ and $\Psi_{ij}$ both as energy-momentum tensor}:
\begin{equation}
\langle \Psi_{ij}(\eta_x,\bm{k})\Psi_{ij}^*(\eta_y,\bm{q})\rangle:=(2\pi)^3\delta^{(3)}(\bm{k}-\bm{q})\Psi(\eta_x,\eta_y,k)
\label{eq:neqcor}
\end{equation}
we can expand the left hand side of Eq.(\ref{eq:neqcor}) as:
\begin{equation}
\langle \Psi_{ij}(\eta_x,\bm{k})\Psi_{ij}^*(\eta_y,\bm{q})\rangle=\int \text{d}^3X\text{d}^3r e^{i(\bm{k}-\bm{q})\cdot\bm{X}}e^{i(\bm{k}+\bm{q})\cdot\frac{\bm{r}}{2}}\langle\Psi_{ij}(\eta_x,\bm{x})\Psi_{ij}(\eta_y,\bm{y})\rangle 
\end{equation}
Where we have defined $\bm{X}:=(\bm{x}+\bm{y})/2$ and $\bm{r}:=\bm{x}-\bm{y}$ to simplify our derivation. Note that the correlator itself doesn't rely on $\bm{x}$ or $\bm{y}$, on the contrary, it only depends on $(\bm{x}-\bm{y})\equiv\bm{r}$. So we have
\begin{equation}
    \text{LHS}=(2\pi)^3\delta^{(3)}(\bm{k}-\bm{q})\int \text{d}^3re^{i\bm{k}\cdot\bm{r}}\langle \Psi_{ij}(\eta_x,\bm{x})\Psi_{ij}(\eta_y,\bm{y})\rangle
\end{equation}
which tells us the factor $\Psi(\eta_x,\eta_y,k)$ appeared in Eq.(\ref{eq:neqcor}) can be written as:
\begin{equation}
    \Psi(\eta_x,\eta_y,k)=\int \text{d}^3re^{i\bm{k}\cdot\bm{r}}\langle \Psi_{ij}(\eta_x,\bm{x})\Psi_{ij}(\eta_y,\bm{y})\rangle
    \label{eq:Psi}
\end{equation}
We know that the energy density of GW is given by\cite{MTW}:
\begin{equation}
    \rho_{\text{GW}}(t/\eta)=\frac{\langle \dot{h}_{ij}(t,\bm{x})\dot{h}_{ij}(t,\bm{x})\rangle}{32\pi G}=\frac{\langle h_{ij}^\prime(\eta,\bm{x}) h_{ij}^\prime(\eta,\bm{x})\rangle}{32\pi Ga^2(\eta)}
    \label{eq:energydensity}
\end{equation}
Note that the $t/\eta$ above means $t$ or $\eta$ rather than $t$ by $\eta$, and here $h_{ij}$ are all transverse and traceless. Now insert Eq.(\ref{eq:solution}) and Eq.(\ref{eq:ab}) into Eq.(\ref{eq:eqcor}) and make an inverse Fourier transformation. Substitute the numerator of Eq.(\ref{eq:energydensity}) by the result we've got, finally we can arrive at:
\begin{equation}
    \rho_{\text{GW}}(\eta)=\frac{2 G\chi^6}{\pi a^4(\eta)}\int \frac{\text{d}k}{k^6}\int_{\zeta_{i}}^{\zeta_{f}}\int_{\zeta_{i}}^{\zeta_{f}}\text{d}\psi\text{d}\phi\ \psi^3\phi^3\cos(\psi-\phi)\Psi(\psi,\phi,k)
\label{eq:rhogw}
\end{equation}
Now we define the dimensionless energy density per logarithmic comoving wave number of GW by:
\begin{equation}
    \Omega_{\text{GW}}(
    \eta,k):=\frac{1}{\rho_{\text{tot}}}\frac{\text{d}\rho_{\text{GW}}}{\text{d}\log k}
    \label{eq:OmegaGW}
\end{equation}
The definition of $\rho_{\text{tot}}$ is unchanged comparing with the definition given in Sec.\ref{sec:2}. Since we consider the spatial curvature equals to zero, the total energy of the universe specifically equals to the energy needs to close the universe i.e. $\rho_{\text{cr}}=\rho_{\text{tot}}$. We assume that the phase transitions happen during RD era, so we can neglect the contribution of energy density by matter $\rho_{\text{mat}}$ and cosmological constant $\rho_{\Lambda}$ at here. 

Substitute $\rho_{\text{GW}}$ by Eq.(\ref{eq:rhogw}), we can rewrite $\Omega_{\text{GW}}$ as:
\begin{equation}
\Omega_{\text{GW}}(\eta,k)=\kappa^2\Bigg(\frac{\alpha}{1+\alpha}\Bigg)^2\Delta^{\text{F}}(k/\widetilde{\beta};\sigma;\eta)
\end{equation}
Here, we use superscript F to indicate FLRW spacetime. The dimensionless quantity $\sigma:=\frac{\widetilde{\beta}}{\mathcal{H}}=\frac{\beta}{H}$ denotes the fraction of speed of the phase transitions to the expansion speed of the universe as the same with the definition given in the previous section. Define $\tau$ as the effective duration of phase transitions, since $\tau$ itself is dependent on $\sigma$, different values of $\sigma$ will influence the upper limit of integral of $\eta_x$ and $\eta_y$ in Eq.(\ref{eq:Delta}), so $\Delta$ is also a function of $\sigma$. 

At this stage, we must remind readers again that our definition of $\Delta^{\text{F}}$ is a little bit different with the definition of $\Delta$ given by
the first line of Eq.(\ref{eq:defDeltaJinno}). Our $\Delta^{\text{F}}$ is a function of $k/\widetilde{\beta}, \sigma$ and $\eta$, while the $\Delta$ defined by Jinno and Takimoto is a function of $k/\beta$ and the speed of bubble wall $v$. The difference appears because of the different treatments of the duration of the phase transitions. In their previous work, they regard the phase transitions start at $t_i=-\infty$ and end at $t_f=+\infty$ in their calculation, which indeed simplify their analyses and decouple $\sigma$ from $\Delta$, but also could bring about a practical problem in latter numerical integration. In the actual integration, we cannot use $\infty$ as the upper limit, on the contrary, we have to artificially specify a finite value which is large enough to confirm the convergence of the integral. However, if we don't know the specific relationship between $\tau$ and $\sigma$, then we may specify a too large number as upper limit of integration which can definitely lead to a longer time for computation. On the other hand, if we pick a too small value as the upper limit of integration, the result we get will be inaccurate and lack of value for reference. Besides the reasons above, there is another important motivation to consider a finite duration of phase transitions. Guo \textit{et al.} consider a finite life time of the sources in their work to study the GW spectrum from sound waves during the first order cosmological phase transitions in an expanding universe.\cite{Guo_2020} As a result, they find an additional suppression factor $\Upsilon(y)$ should be included in the new expression of GW spectrum which could decrease GW spectrum significantly when the life time of sources is quite short. This point can be easily checked according to the FIG.15 of Ref.\cite{Guo_2020}. Be inspired by their work, we think it is necessary to take the finite duration of phase transitions into account. So in our work, $\tau$ itself is a function of $\sigma$ which induces that $\sigma$ cannot be decoupled from $\Delta^{\text{F}}$ anymore. As a result, the value of $\Delta^{\text{F}}$ in our work cannot compare with $\Delta$ in Ref.\cite{Jinno_2017} directly. In fact $\Delta^{\text{F}}$ and $\Delta^{\text{M}}$ are two quantities we will make a comparison later, where $\Delta^{\text{F}}$ represents our result and $\Delta^{\text{M}}$ represents Jinno's result.  

Using Friedmann Equation\cite{Weinberg}:
\begin{equation}
    \mathcal{H}^2=\frac{8\pi G}{3}\rho_{\text{tot}}a^2(\eta)
\end{equation}
$\Delta^{\text{F}}(k/\widetilde{\beta};\sigma;\eta)$ can be rewritten as:
\begin{equation}
\Delta^{\text{F}}(k/\widetilde{\beta};\sigma;\eta)=\frac{3k^3\mathcal{H}^8(\eta)}{4\pi^2\kappa^2\rho_0^2}\int_{\eta_{i}}^{\eta_{f}}\int_{\eta_{i}}^{\eta_{f}}\text{d}\eta_x\text{d}\eta_y\times\Bigg(\eta_x^3\eta_y^3\cos[k(\eta_x-\eta_y)]\Psi(\eta_x,\eta_y,k)\Bigg)
\label{eq:Delta}
\end{equation}
Now, by comparing our Eq.(\ref{eq:Delta}) with Eq.(\ref{eq:Delta_jinno}), we can easily find the difference between these two equations out of two different choices of spacetime background. We'll see the difference more clearly later from Eq.(\ref{eq:Deltas}) and Eq.(\ref{eq:Deltad}). At here, we adopt the Eq.(22) of Ref.\cite{Kosowsky_1994} to describe the functional relationship between $\kappa$ and $\alpha$ for latter numerical estimation as a benchmark: 
\begin{equation}
    \kappa=\frac{1}{1+0.715\alpha}\Bigg[0.715\alpha+\frac{4}{27}\sqrt{\frac{3\alpha}{2}}\Bigg]
    \label{eq:relation}
\end{equation}
Note that this equation only holds for the case of Jouguet detonation, readers can adopt the expressions given by Ref.\cite{Espinosa_2010} to do more estimations when $\alpha$ and $v$ take different values. In Ref.\cite{Espinosa_2010}, Espinosa \textit{et al.} studied all of the bubble expansion regimes without specifying any particle physics model and gave the fitting formulas of $\kappa(v, \alpha)$ in their Appendix A. Note that since our derivation is only valid for phase transitions happen during the RD era, so we artificially restrict $\alpha\in[0,1]$. The work considering other eras and other ranges of $\alpha$ needs to be done in the future.

Now we can clearly discover that the most important quantity we need to focus on is $\Delta^{\text{F}}(k/\widetilde{\beta};\sigma;\eta)$. The coefficients $\kappa$ and $\alpha$ should be uniquely specified by choosing a specific model. In our work, we don't choose any specific model and regard $\alpha$ as a free variable, $\kappa$ is a function of $\alpha$ described by Eq.(\ref{eq:relation}). Through this approach, we can obtain a general understanding of the behavior of GW spectrum without being restricted by any specific model.

Note that the $\Omega_{\text{GW}}$ given by Eq.(\ref{eq:OmegaGW}) denotes the dimensionless energy density of GW per logarithmic comoving wave number when GW has just been generated after the phase transitions rather than nowadays. Considering the effect of cosmological redshift yields:
\begin{equation}
    h_0^2\Omega_{\text{GW}}=1.60\times10^{-5}\kappa^2\Bigg(\frac{g_\star}{106.75}\Bigg)^{-1/3}\Bigg(\frac{\alpha}{1+\alpha}\Bigg)^2\Delta^{\text{F}}(k/\widetilde{\beta};\sigma;\eta)
\label{eq:nowadays}
\end{equation}
\begin{equation}
\begin{aligned}
f&=\frac{k}{2\pi a_0}=\frac{1}{2\pi}\Bigg(\frac{k}{\widetilde{\beta}}\Bigg)\Bigg(\frac{\widetilde{\beta}}{\mathcal{H}}\Bigg)\Bigg(\frac{a_\star}{a_0}\Bigg)H_\star\\
&=2.63\times 10^{-6}\text{Hz}\Bigg(\frac{k}{\widetilde{\beta}}\Bigg)\Bigg(\frac{\widetilde{\beta}}{\mathcal{H}}\Bigg)\Bigg(\frac{T_\star}{100\text{GeV}}\Bigg)\Bigg(\frac{g_\star}{106.75}\Bigg)^{1/6}
\end{aligned}
\label{eq:frequency}
\end{equation}
where $k$ is the magnitude of comoving wave vector $\bm{k}$, $T_\star$ is the temperature after the phase transitions and $g_\star$ is the relativistic degrees of freedom in the universe corresponding to the temperature  $T_\star$.
\subsection{Notations}
For the simplicity of comparison, we adopt the notation system defined in Ref.\cite{Jinno_2017}. The only difference are listed below. We define two new time variables $\Xi$ and $\xi$ rather than the choice of $T$ and $t_\text{d}$:
\begin{equation}
    \Xi:=\frac{\eta_x+\eta_y}{2}\quad\xi:=\eta_x-\eta_y
\end{equation}
Besides these two variables, we only need to replace $t$ with $\eta$ in our derivation. The specific physical meaning of these quantities can refer to the FIG.3 and FIG.5 of Ref.\cite{Jinno_2017}. Out of the special property of FLRW metric, we find that light still move along the 45 degree line in the $\eta-x$ plane of the spacetime diagram, which indicates that we can still use the physical picture given by Jinno and Takimoto directly and make our analyses easier to follow.
\section{\label{sec:4}Analysis and derivation}
In this section, we begin to derive the analytical expressions of $\Delta^{\text{F}(s)}$ and $\Delta^{\text{F}(d)}$ by making use of the method proposed by Jinno and Takimoto.

We define the phase transitions start when the probability for a bubble to nucleate in a Hubble volume during a Hubble time reaches $\mathcal{O}(1)$. Our definition of the end of phase transitions is a little bit tricky which will be shown later. Firstly, let's consider the information that can be derived from our definition of the moment when the phase transitions start. According to the definition, we have:
\begin{equation}
    \frac{\Gamma_\star}{\mathcal{H}_\star^4}\simeq1
\end{equation}
Without loss of generality, we can take the approximate equal sign as equal sign at here. So we have
\begin{equation}
    \Gamma_\star=\mathcal{H}_\star^{4}=\frac{1}{\sigma^4}
\end{equation}
where we have set $\widetilde{\beta}=1$. Besides this, we know that the Hubble parameter related with scale factor during RD era by $H(a)=\chi/a^2$. So we can find that
\begin{equation}
    \widetilde{\beta}=a_\star\beta=a_\star\sigma H_\star=\frac{\sigma\chi}{a_\star}=\frac{\sigma}{\eta_\star}\equiv 1
\end{equation}
Now we have found that the conformal time corresponding to the start of the phase transitions is $\eta_\star=\sigma$. Insert $\eta_\star=\sigma, \widetilde{\beta}=1$ and $\Gamma_\star=1/\sigma^4$ into Eq.(\ref{eq:nucleate}), we have
\begin{equation}
    \Gamma(\eta)=\frac{1}{\sigma^4}e^{\eta-\sigma}
\end{equation}
At this stage, we could see the advantage to adopt FLRW spacetime is twofold. On the one hand, we take a more realistic spacetime background which can reflect the expansion of the universe and then directly influence the final GW spectrum. On the other hand, we have a relatively natural way to connect $\eta$ and $\mathcal{H}$ which could help us determine the start point of phase transitions and therefore we can try to figure out the duration of phase transitions rather than take $\eta_i=-\infty$ and $\eta_f=+\infty$ or manually specify a cutoff which might bring about errors or lead to a longer integration time.

To prepare for the latter calculation, we need to study the false vacuum probability at first.
\subsection{False vacuum probability}
According to Eq.(42) and Eq.(43) of Ref.\cite{Jinno_2017}, we have the probability of two spacetime points staying in the false vacuum state as:
\begin{equation}
    P(x,y)=e^{-I(x,y)}\quad I(x,y)=\int_{V_{xy}}\text{d}^4z\Gamma(z)
\end{equation}
Where the integral of $I(x,y)$ can be expanded as two terms:
\begin{equation}
\left\{
\begin{aligned}
&I_x^{(y)}=\int_{\sigma}^{\eta_{xy}}\frac{\pi}{3}r_x^3\Gamma(\eta)(2+c_x)(1-c_x)^2\text{d}\eta+\int_{\eta_{xy}}^{\eta_x}\frac{4\pi}{3}r_x^3\Gamma(\eta)\text{d}\eta\\
&I_y^{(x)}=\int_{\sigma}^{\eta_{xy}}\frac{\pi}{3}r_y^3\Gamma(\eta)(2-c_y)(1+c_y)^2\text{d}\eta+\int_{\eta_{xy}}^{\eta_y}\frac{4\pi}{3}r_y^3\Gamma(\eta)\text{d}\eta
\end{aligned}
\right.
\end{equation}
After integration, we have $I(x,y)=I_x^{(y)}+I_y^{(x)}$:
\begin{equation}
\begin{aligned}
 I(x,y)=&\frac{\pi}{12 r \sigma ^4} \Bigg\{ \Bigg[-12\xi ^2 \Bigg(\Xi ^2-2 (\Xi +1) \sigma +2 \Xi +\sigma ^2+2\Bigg)-3 r^2 \Big(\xi ^2+4(\Xi ^2-2 (\Xi +1) \sigma\\
 &+2 \Xi +\sigma ^2+2)\Big)+r^4-4 r \Big(3 \xi ^2
   (\Xi -\sigma +1)+24 (\Xi -
 \sigma +1)+4 (\Xi-\sigma )^2(\Xi -\sigma \\
 &+3)\Big)\Bigg]+24e^{\Xi-\frac{r }{2}-\sigma } \Big(\xi ^2-r (r+4)\Big)+
 48 r e^{\Xi-\sigma}\cosh(\frac{\xi}{2})\Bigg\}
\end{aligned}
\end{equation}

Before we go ahead, let's explain the reason why we need to discuss the probability of spacetime points staying in the false vacuum. Remember that under the so-called envelope approximation, any spatial point can be passed by a bubble for only once. Assuming that a bubble wall passes spatial point $\bm{x}$ at $\eta=\eta_x$, we must confirm that there are no bubbles nucleated inside the past cones of $x$, otherwise the spatial point $\bm{x}$ must have already transformed from the false vacuum state to the true vacuum state before $\eta=\eta_x$ and cannot be passed by another bubble by the second time. For the sake of the reason above, we have to figure out the probability of spacetime points staying in the false vacuum at the very first step of our calculation.

\subsection{Contributions from single bubble case and double bubble case}
So far, we have every ingredient prepared to analyze the energy spectrum of GW. Just as the definition given by Jinno and Takimoto, $\Delta^{\text{F}}$ can be decomposed as $\Delta^{\text{F}(s)}+\Delta^{\text{F}(d)}$, where $\Delta^{\text{F}(s)}$ indicates the contribution to $\Delta^{\text{F}}$ from single bubble case and $\Delta^{\text{F}(d)}$ denotes the contribution from double bubble case. The specific meaning of these two cases have been discussed in Sec.\ref{sec:2}. Here we only simply introduce the calculation procedure taking the single bubble case for an example and
directly give our results, the detailed calculation ideas can refer to the Appendix A of Ref.\cite{Jinno_2017}. Remember that our ultimate target is to get the GW spectrum which can be calculated by Eq.(\ref{eq:OmegaGW}) and Eq.(\ref{eq:Delta}). In order to use these two equations we must figure out $\Psi(\eta_x,\eta_y,k)$ defined by Eq.(\ref{eq:Psi}) at first, which requires us to find all of the possible spacetime points for bubbles to nucleate which can lead to non-vanishing energy-momentum tensor at two spacetime points $x$ and $y$ to make $\langle \Psi_{ij}(x)\Psi_{ij}(y)\rangle$ nonzero. When we consider the single bubble case, we'll easily find that only if the bubble nucleates at some special points which are belonging to the intersection part of the past cones of $x$ and $y$ (in Ref.\cite{Jinno_2017}, they call it $\delta V_{xy}$), may this situation contribute to $\Delta^{\text{F}(s)}$. By taking the possibility for bubbles to nucleate in these spacetime points into consideration and calculating the contributions given by every single configuration, we can write the two-point correlator of energy-momentum tensor as below:

\begin{equation}
\langle \Psi_{ij}(\eta_x,\bm{x})\Psi_{ij}(\eta_y,\bm{y})\rangle^{(s)} =\frac{2\pi}{9r^5\sigma^4}\kappa^2\rho_0^2P(\eta_x,\eta_y,r)\times\Bigg[\frac{1}{2}F_0+\frac{1}{4}(1-c_{rk}^2)F_1+\frac{1}{16}(1-c_{rk}^2)F_2\Bigg]
\end{equation}
where $F_0, F_1$ and $F_2$ are given by:
\begin{equation}
\begin{aligned} 
F_0=&-\frac{(r^2-\xi^2 )^2 e^{-\frac{r}{2}-\sigma }}{16}
   \Bigg\{e^{\frac{r}{2}+\sigma } \Bigg[16 (\Xi -\sigma )^2 \left(\Xi ^2-2 (\Xi
   +2) \sigma +4 \Xi +\sigma ^2+12\right)+384 (\Xi -\sigma +1)-\\
   &8 r^2 (\Xi^2-2 (\Xi +1) \sigma+2 \Xi +\sigma ^2+2)+r^4\Bigg]-32 e^{\Xi } (r(r+6)+12)\Bigg\}
\end{aligned}
\end{equation}
\begin{equation}
\begin{aligned} 
F_1=&-\frac{(r^2-\xi^2 )e^{-\frac{r}{2}-\sigma }}{8}\Bigg\{e^{\frac{r}{2}+\sigma }
   \Bigg[(80 \xi ^2 (\Xi -\sigma ) \Big[\Xi ^2 (4-3 \sigma )+\Xi ^3+\Xi
   (\sigma  (3 \sigma -8)+12)-\sigma  ((\sigma -4) \sigma\\
   &+12)+24\Big]+1920\xi ^2+r^4 (\xi ^2-8 (\Xi ^2-2 (\Xi +1) \sigma+2 \Xi+\sigma^2+2))+8 r^2 \Big(-3 \xi ^2 (\Xi ^2-2 (\Xi\\
   &+1) \sigma +2\Xi+\sigma ^2+2)-2 (\Xi -\sigma )^2 (\Xi ^2-2 (\Xi +2) \sigma+4\Xi +\sigma ^2+12)-48 (\Xi -\sigma +1)\Big)+3 r^6\Bigg]\\
   &+16 e^{\Xi }(r^2 (r\times(r (r+4)+12)+24)-\xi ^2 (r (r (r+12)+60)+120))\Bigg\}
\end{aligned}
\end{equation}
\begin{equation}
\begin{aligned}
\nonumber F_2=&-\frac{1}{16}\Bigg\{560 \xi ^4 (\Xi ^4+4 \Xi ^3+12 \Xi ^2-4 (\Xi +1) \sigma ^3+6 (\Xi 
   (\Xi +2)+2) \sigma ^2-4 (\Xi  (\Xi(\Xi +3)+6)+6) \sigma\\
   &+24 \Xi+\sigma
   ^4+24)+2 r^6 (\xi ^2+4 (\Xi ^2-2 (\Xi +1) \sigma +2 \Xi
   +\sigma ^2+2))+3 r^4 (16 \xi ^2 (\Xi ^2-2 (\Xi +1)
   \sigma \\
   &+2 \Xi +\sigma ^2+2)+\xi ^4+16 (\Xi -\sigma )^2 (\Xi ^2-2(\Xi+2) \sigma +4 \Xi +\sigma ^2+12)+384 (\Xi -\sigma +1))\\
   &-120\xi ^2 r^2 (\xi ^2 \Xi ^2-2 (\Xi +1) \sigma +2 \Xi+\sigma^2+2)+4 (\Xi -\sigma )^2 (\Xi ^2-2 (\Xi +2) \sigma +4 \Xi +\sigma
   ^2+12)\\
   &+96 (\Xi -\sigma +1))+(16 \xi ^2 (r (r (r
   (r+12)+84)+360)+720) r^2-8 (r (r (r (r+4)+20)+72)\\
   &+144) r^4-8 \xi ^4 (r (r (r(r+20)+180)+840)+1680))e^{\Xi -\frac{r}{2}-\sigma }+3 r^8\Bigg\}
\end{aligned}
\end{equation}
So $\Psi^{(s)}(\eta_x,\eta_y,k)$ can be rewritten as following:
\begin{equation}
\Psi^{(s)}(\eta_x,\eta_y,k)=\int \text{d}^3re^{i\bm{k}\cdot\bm{r}}\frac{2\pi}{9r^5\sigma^4}\kappa^2\rho_0^2\times P(\eta_x,\eta_y,r)\Bigg[\frac{1}{2}F_0+\frac{1}{4}(1-c_{rk}^2)F_1+\frac{1}{16}(1-c_{rk}^2)^2F_2\Bigg]
\label{eq:Psis}
\end{equation}
Here $c_{rk}$ is the shorthand of $\cos (\langle\hat{\bm{r}},\hat{\bm{k}}\rangle )$, where $\langle\bm{a},\bm{b}\rangle$ denotes the angle between $\bm{a}$ and $\bm{b}$ rather than ensemble average. 

\begin{figure}
    \centering
    \includegraphics[width=0.6\textwidth]{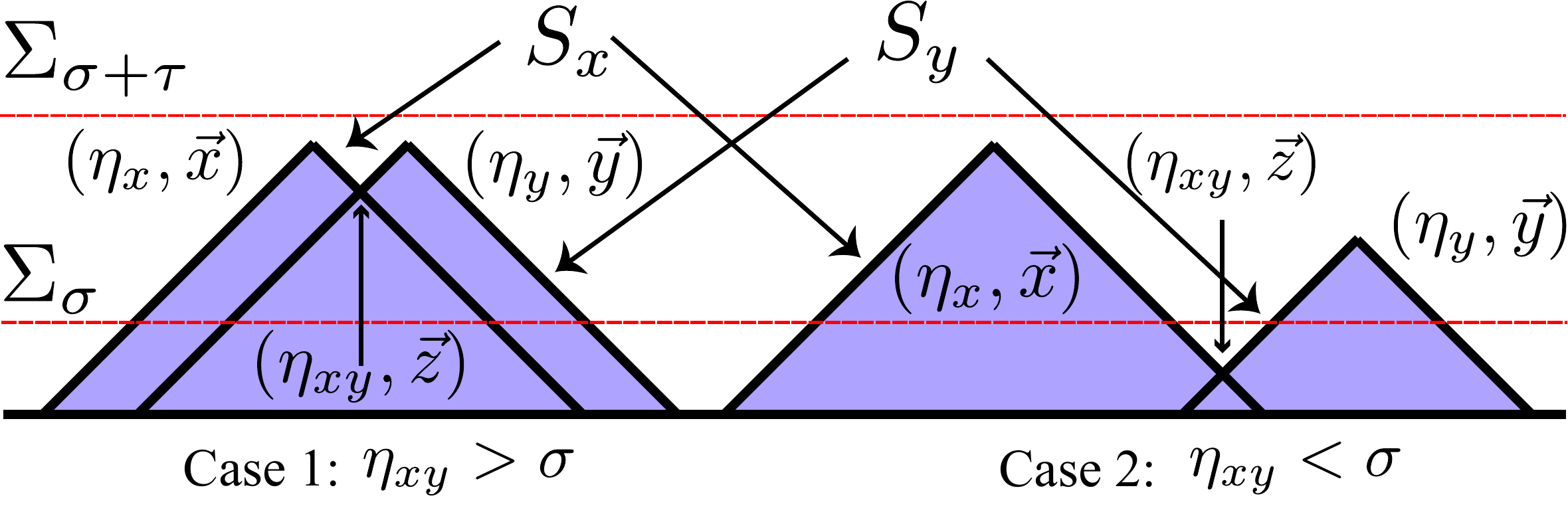}
    \caption{Determination of the upper limit of $r$ in the integral Eq.(\ref{eq:Psis}). $(\eta_x,\vec{x})$ and $(\eta_y,\vec{y})$ denote two spacetime points, $S_x,S_y$ are their past cones, $\Sigma_\sigma$ and $\Sigma_{\sigma+\tau}$ characterize two hyperplanes where $\eta=\sigma$ and $\eta=\sigma+\tau$ respectively. Case 1 corresponds to the situation where $\eta_{xy}>\sigma$, under such a circumstance, the bubble can nucleate after the beginning of the phase transitions and influence two spacetime points $x,y$. Case 2 corresponds to the situation where $\eta_{xy}<\sigma$, so the bubble must nucleate before the start of the phase transitions which is forbidden.}
    \label{fig:determination}
\end{figure}

The integration of $\phi$ is trivial, since all of the terms show up in the integrand are independent with $\phi$. However, the integration of $r$ and $\theta$ need to be paid more attention. Let's consider the integration of $r$ first. The determination of lower and upper limit of the integration of $r$ needs to be careful. Note that the separation of $x$ and $y$ must be space-like, one can easily find that if the separation is time-like, there must exist a spatial point ($\bm{x}$ or $\bm{y}$) which has transformed from the false vacuum state to the true vacuum state before $\eta_x$ or $\eta_y$. According to this requirement, we have $r_{\min}=\abs{\eta_x-\eta_y}=\abs{\xi}$. Without loss of generality, we can assume $\eta_x>\eta_y$, so we are permitted to throw away the absolute sign, and denote $r_{\min}=\xi$ directly. The process of deciding the upper limit of $r$ is a little bit complicated. We can see FIG.\ref{fig:determination} for a straightforward understanding.

According to FIG.\ref{fig:determination}, we can find that only if $\eta_{xy}>\sigma$ can this configuration contributes to the integration, so we have:
\begin{equation}
    \eta_{xy}:=\frac{\eta_x+\eta_y-r}{2}\geqslant\sigma\Longrightarrow r\leqslant2(\Xi-\sigma)
\end{equation}
With the upper and low limits for the integration of $r$ being decided, we can perform integration and rewrite $\Psi^{(s)}(\eta_x,\eta_y,k)$ as following:
\begin{equation}
\Psi^{(s)}(\eta_x,\eta_y,k)=\int_{\xi}^{2(\Xi-\sigma)}\text{d}r\ \frac{4\pi^2\kappa^2\rho_0^2}{9r^3\sigma^4}e^{-I(x,y)}\times\Bigg[j_0(kr)F_0+\frac{j_1(kr)}{kr}F_1+\frac{j_2(kr)}{(kr)^2}F_2\Bigg]
\end{equation}
where $j_0(kr),j_1(kr),j_2(kr)$ are spherical Bessel functions. Here we have used integration formula to perform the integration of $\theta$:
\begin{equation}
	\int_{-1}^{+1}\text{d}c\  e^{icx}=2j_0(x)\quad \int_{-1}^{+1}\text{d}c\ e^{icx}(1-c^2)=\frac{4j_1(x)}{x}\quad\int_{-1}^{+1}\text{d}c\ e^{icx}(1-c^2)^2=\frac{16j_2(x)}{x^2}
\end{equation}
Recall Eq.(\ref{eq:Delta}) and insert $\Psi^{(s)}(\eta_x,\eta_y,k)$ into it, we have
\begin{equation}
\begin{aligned}
\Delta^{\text{F}(s)}(k/\widetilde{\beta};\sigma;\eta_{f})=\frac{2k^3}{3\sigma^4(\sigma+\tau)^8}\int_\sigma^{\sigma+\tau}\text{d}\Xi&\int_0^\tau\text{d}\xi\int_\xi^{2(\Xi-\sigma)}\text{d}r\times\\
&\Bigg[\frac{e^{-I(x,y)}}{r^3}\Bigg(\Xi^2-\frac{\xi^2}{4}\Bigg)^3\cos(k\xi)\mathcal{J}^{(s)}(k,r,\Xi,\xi,\sigma)\Bigg]
\end{aligned}
\label{eq:Deltas}
\end{equation}
where $\mathcal{J}^{(s)}(k,r,\Xi,\xi,\sigma)$ is defined by:
\begin{equation}
\mathcal{J}^{(s)}(k,r,\Xi,\xi,\sigma)\equiv \Bigg[j_0(kr)F_0+\frac{j_1(kr)}{kr}F_1+\frac{j_2(kr)}{(kr)^2}F_2\Bigg]
\end{equation}
Now we have got the energy spectrum contributed by the single bubble case successfully. We can see that the final expression of $\Delta^{\text{F}(s)}$ can be described by a triple integration, while in Jinno and Takimoto's work $\Delta^{\text{M}(s)}$ is given by a double integration. At here, we can compare Eq.(\ref{eq:Deltas}) with Eq.(54) of Ref.\cite{Jinno_2017} to see the difference brought by the different choices of spacetime background. Neglecting the difference of constant coefficients like $2/3$ and $1/12$, we can find the significant difference between two denominators. Factors $(\sigma+\tau)^8$ and $(\Xi^2-\xi^2/4)^3$ appear because we choose FLRW spacetime and $h_{ij}$ thereby satisfy a different equation of motion which could make an impact to the final result. The factor $\tau$ in $(\sigma+\tau)^8$ especially shows the dilution effect led by the expansion of the universe. Although the collision of bubbles generally emits a bunch of energy, the lapse of time also keeps diluting it. 

Next, we're going to directly show the contribution to the total $\Delta^{\text{F}}$ from double bubble case i.e. $\Delta^{\text{F}(d)}$. Readers can turn to the Part D of Sec.III of Ref.\cite{Jinno_2017}  for the calculation ideas and details .

As the same with the single bubble case, in order to obtain $\Delta^{\text{F}(d)}$, we need to find the expression of $\Psi^{(d)}$:
\begin{equation}
\Psi^{(d)}(\eta_x,\eta_y,k)=16\pi\int_\xi^{2(\Xi-\sigma)}r^2\text{d}rP(\eta_x,\eta_y,r)\times \mathcal{B}_x^{(d)}(\eta_x,\eta_y,r)\mathcal{B}_y^{(d)}(\eta_x,\eta_y,r)\frac{j_2(kr)}{(kr)^{2}}
\end{equation}
where $\mathcal{B}^{(d)}_x(\eta_x,\eta_y,r)$ and $\mathcal{B}^{(d)}_y(\eta_x,\eta_y,r)$ are given by:
\begin{equation}
\begin{aligned}
\mathcal{B}_x^{(d)}(\eta_x,\eta_y,r)=&-\frac{\pi  \kappa  \rho_0  (r^2-\xi^2 )e^{-\frac{r}{2}-\sigma }}{24 r^3 \sigma ^4} (e^{\frac{r}{2}+\sigma } (-8 \xi\times(\Xi -\sigma ) (\Xi ^2+\Xi (3-2 \sigma )+
   (\sigma -3) \sigma\\
   &+6)-48 \xi+2 r^2 (\xi  (\Xi
   -\sigma +1)-2 (\Xi ^2-2 (\Xi +1) \sigma +2 \Xi +\sigma^2+2))+r^4)\\
   &+4 e^{\Xi } (12 \xi +r (6 \xi +r (\xi+r+2))))
\end{aligned}
\end{equation}
\begin{equation}
\begin{aligned}
\mathcal{B}_y^{(d)}(\eta_x,\eta_y,r)=&-\frac{\pi  \kappa  \rho_0  (r^2-\xi^2 ) e^{-\frac{r}{2}-\sigma }}{24 r^3 \sigma ^4}
   (e^{\frac{r}{2}+\sigma } (8 \xi\times(\Xi -\sigma ) (\Xi ^2+\Xi 
   (3-2 \sigma )+(\sigma -3) \sigma+6)\\
   &+48 \xi-2 r^2 (-\sigma  (\xi
   +4 \Xi +4)+\xi  \Xi +\xi +2 \Xi ^2+4 \Xi +2 \sigma ^2+4)+r^4)+4e^{\Xi } \\
   &\times(r^2 (r+2)-\xi  (r (r+6)+12)))
\end{aligned}
\end{equation}
Rearrange the equations above, we'll find
\begin{equation}
\Psi(\eta_x,\eta_y,k)^{(d)}=\int_\xi^{2(\Xi-\sigma)}\text{d}r\ \frac{16\pi^3\kappa^2\rho_0^2}{r^4\sigma^8}P(\eta_x,\eta_y,r)\mathcal{J}^{(d)}(k,r,\Xi,\xi,\sigma)
\label{eq:Psid}
\end{equation}
where we have defined three new quantities whose definitions are given below:
\begin{equation}
\mathcal{B}^{(d)}_{x/y}=\frac{\pi\kappa\rho_0}{r^3\sigma^4}\widetilde{\mathcal{B}}^{(d)}_{x/y}\quad \mathcal{J}^{(d)}(k,r,\Xi,\xi,\sigma)=\widetilde{\mathcal{B}}_x^{(d)}(\eta_x,\eta_y,r)\widetilde{\mathcal{B}}_y^{(d)}(\eta_x,\eta_y,r)\frac{j_2(kr)}{(kr)^{2}}
\label{eq:Jd}
\end{equation}
Insert Eq.(\ref{eq:Psid}-\ref{eq:Jd}) into Eq.(\ref{eq:Delta}) and then we have the final form of $\Delta^{(d)}(k/\widetilde{\beta};\sigma;\eta_f)$ given by Eq.(\ref{eq:Deltad}):
\begin{equation}
\begin{aligned}
\Delta^{\text{F}(d)}(k/\widetilde{\beta};\sigma;\eta_{f})=\frac{24\pi k^3}{\sigma^8(\sigma+\tau)^8}\int_\sigma^{\sigma+\tau}\text{d}\Xi&\int_0^\tau\text{d}\xi\int_\xi^{2(\Xi-\sigma)}\text{d}r\times\\
&\Bigg[\Bigg(\Xi^2-\frac{\xi^2}{4}\Bigg)^3\cos(k\xi)\frac{e^{-I(x,y)}}{r^4}\mathcal{J}^{(d)}(k,r,\Xi,\xi,\sigma)\Bigg]
\end{aligned}
\label{eq:Deltad}
\end{equation}
Adding $\Delta^{\text{F}(s)}$ and $\Delta^{\text{F}(d)}$ up, we can get the analytical expression of $\Delta^{\text{F}}$, which is specifically the  quantity we want to obtain. At this step, we have no idea about how to go forward by analysis anymore. The analytical expressions of two triple integrations given by Eq.(\ref{eq:Psis}) and Eq.(\ref{eq:Psid}) are too complicated that we cannot integrate them directly. However, we can integrate them numerically to see the behavior of $\Delta^{\text{F}}$ versus different values of $\sigma$. We'll show the result in the next section and make comparison between our results with the results in Minkowski spacetime.  

So far we have successfully obtained the analytical expressions of $\Delta^{\text{F}}$, but how can we know the validity of our derivation? One of the good approaches is to consider a limiting case: Does our results converge to the results given in Minkowski spacetime when $\sigma\to\infty$ ? When $\sigma\to\infty$, we have $\tau/\sigma\to0$ which means that the impact brought by the expansion of the universe disappears, as a result we don't need to consider it anymore. So in principle, our results are supposed to converge to the expressions given in Minkowski spacetime in the limiting case of large $\sigma$. We can prove it in two steps. Firstly, let's recall Eq.(\ref{eq:Delta}) and compare it with Eq.(\ref{eq:Delta_jinno}). We can find that if we ignore the difference between $\Psi(\eta_x,\eta_y,k)$ and  $\Pi(t_x,t_y,k)$, then the only difference between two integrations is that $\Delta^\text{F}$ has an additional factor $\mathcal{H}^8\eta_x^3\eta_y^3$. By derivation, we have:

\begin{equation}
      \frac{\sigma^3\sigma^3}{(\sigma+\tau)^8}\leqslant\mathcal{H}^8(\eta_f)\eta_x^3\eta_y^3\leqslant\frac{(\sigma+\tau)^3(\sigma+\tau)^3}{(\sigma+\tau)^8}
    \label{eq:limit}
\end{equation}
When $\sigma\to\infty$, the phase transitions finish instantly , as a result $\tau/\sigma\to0$, so we can rewrite Eq.(\ref{eq:limit}) as:
\begin{equation}
    \mathcal{H}^8(\eta_f)\eta_x^3\eta_y^3\simeq\frac{1}{\sigma^2}\quad\text{(for large $\sigma$)}
\end{equation}
At this stage, we have successfully extracted the factor $1/\sigma^2$ out of the expression of $\Delta^\text{F}$. Recall that in Eq.(\ref{eq:defDeltaJinno}) we specially define $\Delta^\text{M}$ by dividing $\Delta$ with $\sigma^2$ to make the definition of $\Delta^{\text{M}}$ and $\Delta^{\text{F}}$ consistent with each other. Now the only integrand left is $\cos[k(\eta_x-\eta_y)]\Psi(\eta_x,\eta_y,k)$ which looks really similar to $\cos[k(t_x-t_y)]\Pi(t_x,t_y,k)$ and the next step we need to do is to prove $\Psi(\eta_x,\eta_y,k)\to\Pi(t_x,t_y,k)$ when $\sigma\to\infty$. However, don't forget a very important point, in our derivation the duration of phase transitions $\tau$ is finite while in Jinno and Takimoto's previous paper, they regard it as infinity to simplify their derivation. When $\sigma\to\infty$, we will also have $\tau\to\infty$ although these two infinity are not the same order. So we can naturally set the upper limit of integral $\eta_f=\sigma+\tau$ as $\infty$. The lower limit also can be set as $-\infty$ because when $\eta<\sigma$, there doesn't exist the collision of bubbles, so they won't contribute to the integral. So we can set $\eta_i=-\infty$, $\eta_f=+\infty$ and ignore the difference of $t$ and $\eta$, actually we'll obtain the totally same expressions of $I(x,y), F_0, F_1$ and $F_2$ as in Ref.\cite{Jinno_2017}, thereby leading to the same expressions of  $\Psi(\eta_x,\eta_y,k)$ and $\Pi(t_x,t_y,k)$. Since $t$ and $\eta$ are both integration variables, $\displaystyle{\int_{-\infty}^{+\infty}\int_{-\infty}^{+\infty}\cos[k(\eta_x-\eta_y)]\Psi(\eta_x,\eta_y,k)\text{d}\eta_x\text{d}\eta_y}$ certainly equals to $\displaystyle{\int_{-\infty}^{+\infty}\int_{-\infty}^{+\infty}\cos[k(t_x-t_y)]\Pi(t_x,t_y,k)\text{d}t_x\text{d}t_y}$. Now we can definitely confirm that our results indeed converge to the results given in Minkowski spacetime by taking large $\sigma$. Later, we'll directly give the numerical simulation to show this point.

From the instructions above, actually we can clearly see that the difference of the final results of GW spectrum are contributed by two sources. The first one is the different choices of spacetime metric: our derivation takes the expansion of the universe into consideration, the factor $\mathcal{H}^8(\eta_f)\eta_x^3\eta_y^3$ in integrand shows up for this reason. When $\sigma\sim \mathcal{O}(1)$ this term can play an important role and dilute the energy density of GW. The second source is the different treatments of phase transitions duration. Intuitively, the approximation of infinite duration works worse when $\sigma$ is really small, so to consider a more realistic length of phase transitions is important.

\section{\label{sec:5}Numerical Calculation}
In the previous section, we have obtained the final form of $\Delta^{\text{F}(s)}$ and $\Delta^{\text{F}(d)}$ given by Eq.(\ref{eq:Deltas}) and Eq.(\ref{eq:Deltad}). However, we have not discussed the way to get the “duration” of phase transitions. Now, let's observe the Eq.(\ref{eq:Deltas}) and Eq.(\ref{eq:Deltad}), there is a special  factor$(\sigma+\tau)^8$ in the denominator of both expressions of $\Delta^{\text{F}(s)}$ and $\Delta^{\text{F}(d)}$. This factor coming from $\mathcal{H}^8(\eta_f\equiv\sigma+\tau)$ directly shows the influence of the expansion of the universe. Although the collision of bubbles emits energy, the expansion of the universe dilute it all the time. In another perspective, we can rewrite $\mathcal{H}^8(\sigma+\tau)$ as $1/[\mathcal{H}^{-1}(\sigma+\tau)]^8$ where $\mathcal{H}^{-1}(\sigma+\tau)$ can be understood as the Hubble horizon at time $\sigma+\tau$. As time elapses, the Hubble horizon keeps growing while the released energy from bubble collisions won't increase all the time, as a result, the energy density of GW from bubble collisions in this process will increase first and decrease later. If we only pay attention to the triple integrations in the Eq.(\ref{eq:Deltas}) and Eq.(\ref{eq:Deltad}), we can find that two integrals increase first and converge to specific values at last with increasing $\tau$, nevertheless the prefactor decreases monotonously in this process. So we can expect the value of $\Delta^{\text{F}}=\Delta^{\text{F}(s)}+\Delta^{\text{F}(d)}$ will increase first and decrease after it reaching the maximum. Before the maximum point, the new released energy is more powerful than the dilution effect of the expansion of the universe, so we can regard that the phase transition has not finished. After the maximum point, the new released energy is not enough to compensate for the dilution effect of the expansion of the universe, so we can regard that the phase transition has finished. Now, let's call $\tau$ as “effective duration” of phase transitions. For a given $\sigma$, we need to study the behavior of $\Delta^{\text{F}}$ first to figure out the exact value of effective duration $\tau$ and then to calculate the corresponding energy density. In a word, we define $\tau$ by the following equation:
\begin{equation}
    \tau:=\underset{\tau}{\text{argmax}}\Bigg(\Delta^{\text{F}(s)}+\Delta^{\text{F}(d)}\Bigg)
\end{equation}

After numerically finding the exact value of $\tau$, we can integrate  $\Delta^{\text{F}(s)}$ and  $\Delta^{\text{F}(d)}$ numerically to study the concrete dependence of $\Delta^{\text{F}}$ on $\sigma$. At the same time, we are supposed to compare our results with the results given by Ref.\cite{Jinno_2017} to estimate the influence brought by the differences between two background spacetime and finite duration.
\begin{figure}
    \centering
    \includegraphics[width=1.0\textwidth]{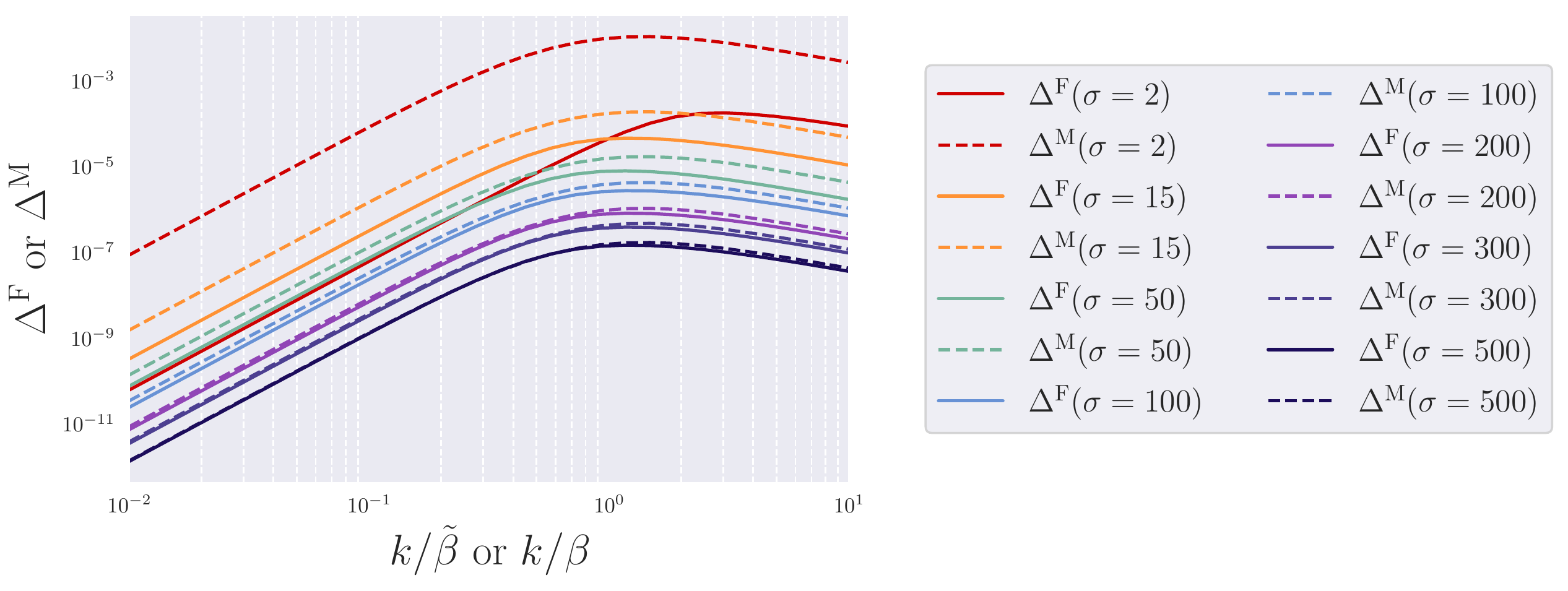}
    \caption{Plot of GW spectrum. The solid curves indicate our results of $\Delta^{\text{F}}$ in FLRW spacetime, while the dashed curves indicates Jinno and Takimoto's results of $\Delta^{\text{M}}$ in Minkowski spacetime. When $\sigma=2\sim\mathcal{O}(1)$, two curves deviate a lot. With increasing $\sigma$, two curves become more and more close. When $\sigma\gtrsim200$, two curves almost coincide with each other as our expectation.}
    \label{fig:compare}
\end{figure}

Firstly, a comparison between our results in FLRW spacetime with a finite duration and results in Minkowski spacetime with an infinite duration is given by FIG.\ref{fig:compare}. Here, we have shown 7 pairs of curves where solid curves indicate our results of $\Delta^{\text{F}}$ estimated by Eq.(\ref{eq:Deltas}) and Eq.(\ref{eq:Deltad}) in FLRW spacetime while dashed curves indicate Jinno and Takimoto's results of $\Delta^{\text{M}}$ estimated by Eq.(\ref{eq:jinnofitting}) in Minkowski spacetime. It's apparent that when $\sigma$ is relative small i.e. we cannot neglect the expansion of the universe, the dashed curves and solid curves deviate a lot. When $\sigma$ generally increase, we can also find the deviation between the dashed curves and solid curves decreases as our expectation. When $\sigma\gtrsim200$, two curves almost coincide with each other. 

\begin{figure}
    \centering
    \includegraphics[width=0.6\textwidth]{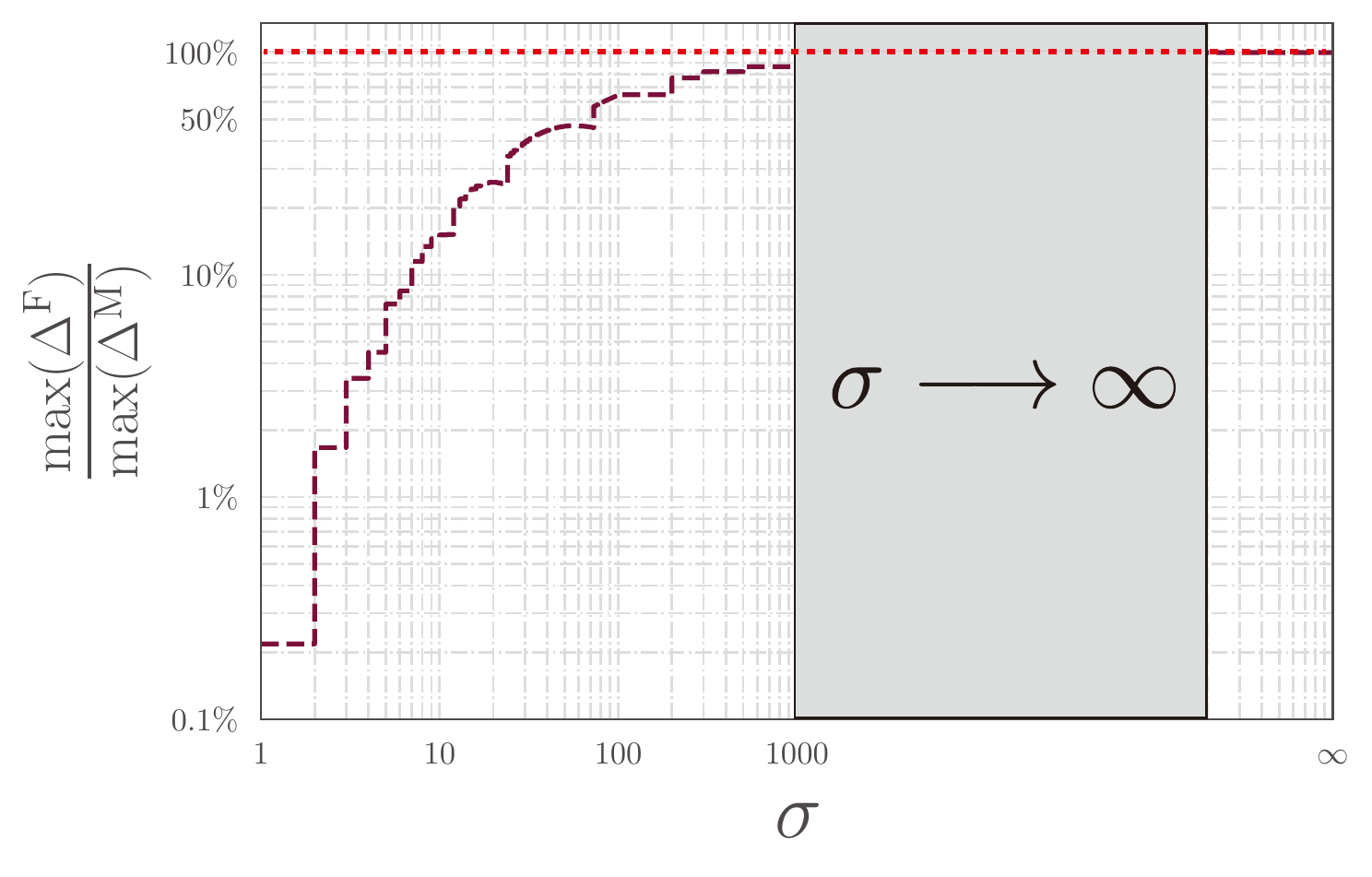}
    \caption{The step plot of the fraction of the maximum value of $\Delta^{\text{F}}$ to the maximum value of $\Delta^{\text{M}}$ versus $\sigma$. When $\sigma\lesssim\mathcal{O}(10)$, the corresponding GW spectrum is significantly influenced by the expansion of the universe. Even when $\sigma\sim100$, the GW spectrum is still be depressed by 50$\%$ compared to the results estimated in Minkowski spacetime. When $\sigma$ increases, this ratio increase as well and will finally converge to $100\%$ when $\sigma\to\infty$, in which regime the expansion of the universe can be totally neglected.}
    \label{fig:ratio}
\end{figure}

According to our numerical estimation, we obtain the behavior of the quantity $\max(\Delta^{\text{F}})$ over $\max(\Delta^{\text{M}})$ versus $\sigma$ shown by FIG.\ref{fig:ratio}. Noticing that with the guarantee of Eq.(\ref{eq:nowadays}), $\Delta^{\text{F}}$ is proportional to $h_0^2\Omega_{\text{GW}}$, so
the significant decrease in $\Delta^{\text{F}}$ can directly lead to weaker signals for detection. Observing FIG.\ref{fig:ratio}, we can clearly find the trend of this curve is the same as our previous expectation. When $\sigma$ is small i.e. the expansion speed of the universe is comparable to the speed of phase transitions, the corresponding GW spectrum is significantly influenced. When $\sigma=1$ a.k.a. $\widetilde{\beta}=\mathcal{H}$, the ratio is only equal to 0.2$\%$. Even when $\sigma=100$, this ratio is still as small as $64.5\%$ which shows the big impact brought by the expansion of the universe and also the consideration of finite duration of phase transitions. When $\sigma$ keeps increasing, this ratio will also increase monotonously and finally converge to 100$\%$ when $\sigma\to\infty$. As our analytical derivation in the end of the last section, when $\sigma\to\infty$, our results should return to the results estimated in Minkowski spacetime and our numerical calculation successfully 
shows this point.

\begin{figure*}
\includegraphics[width=0.5\textwidth]{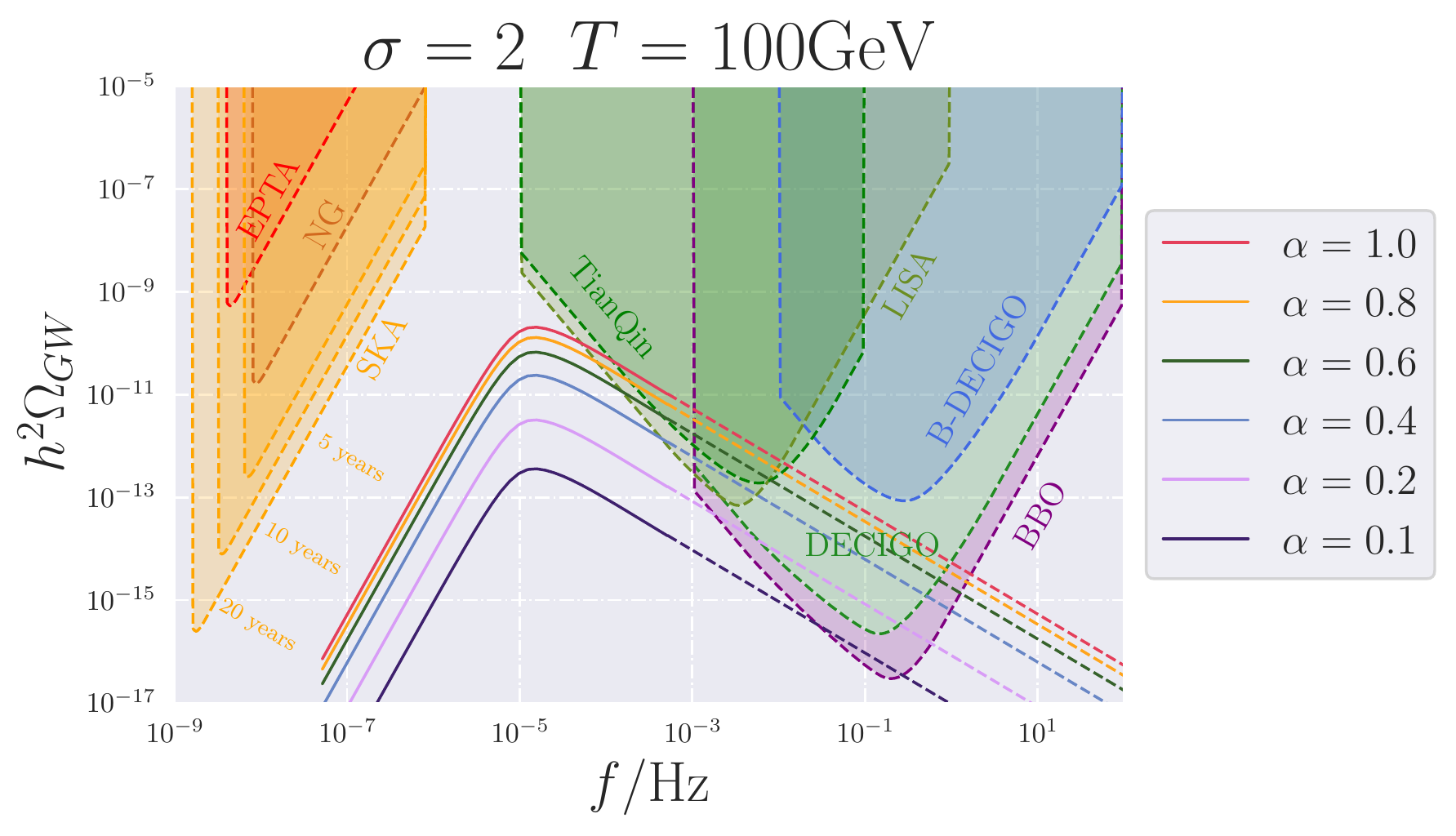}
\includegraphics[width=0.5\textwidth]{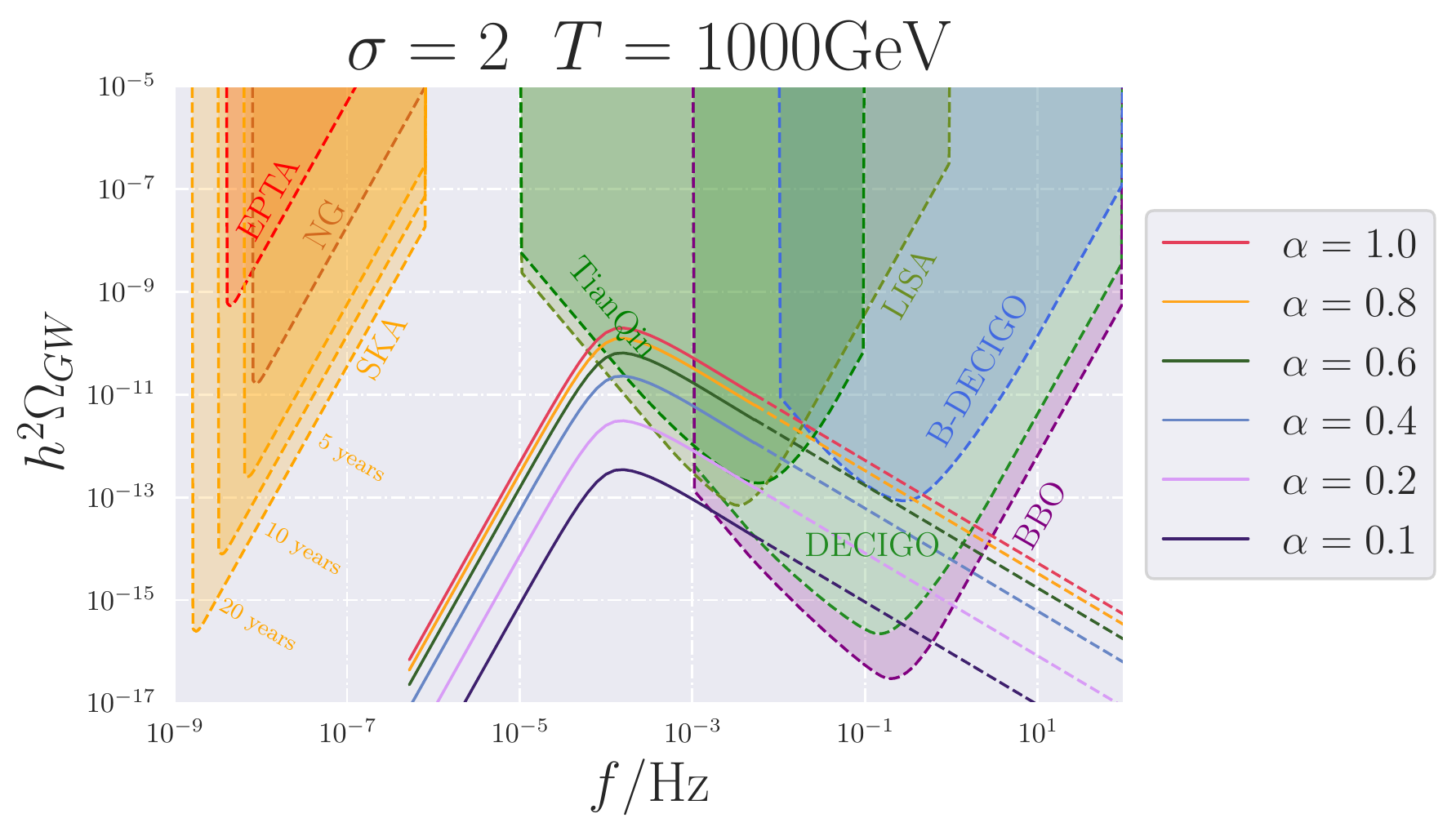}
\includegraphics[width=0.5\textwidth]{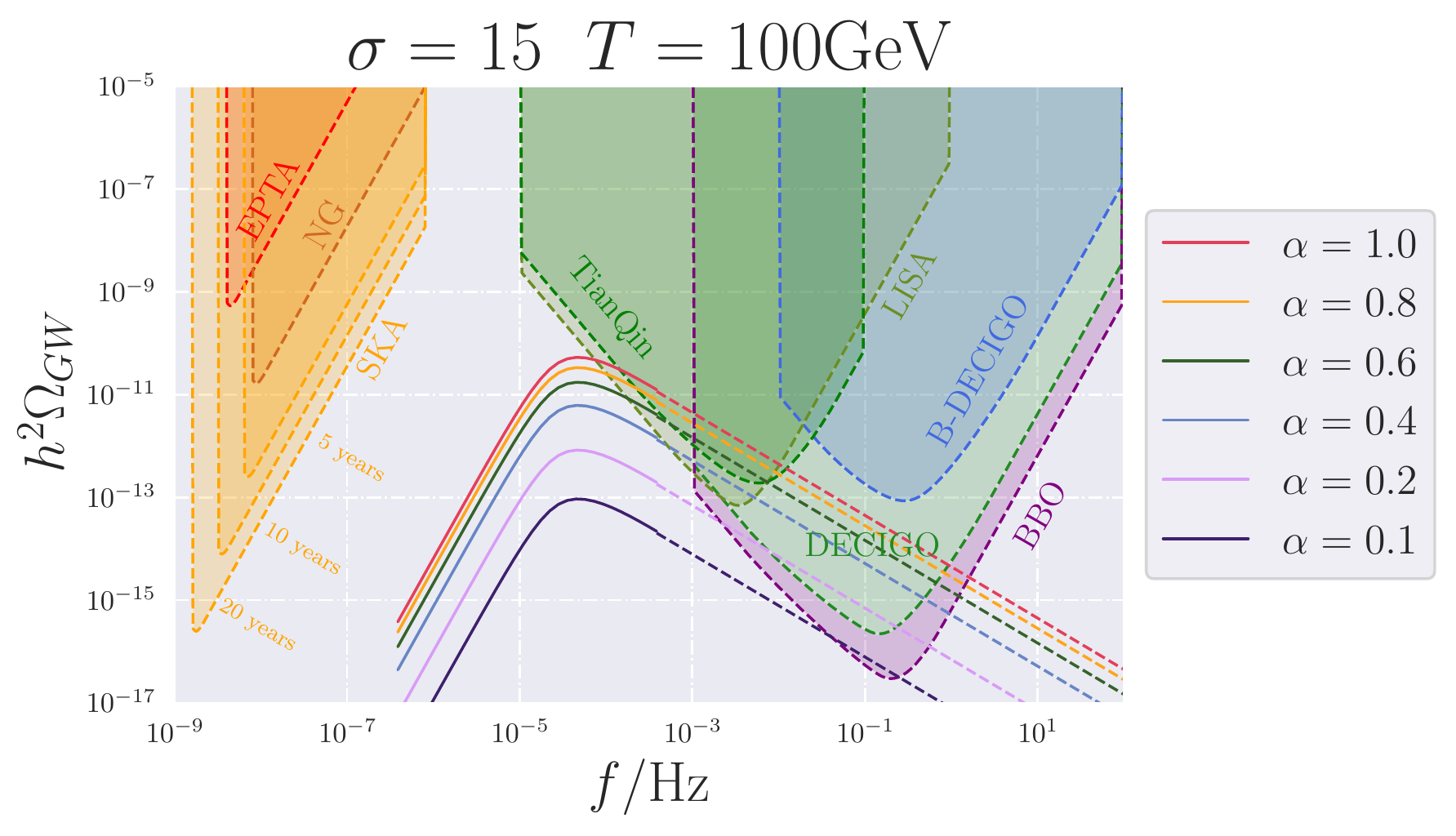}
\includegraphics[width=0.5\textwidth]{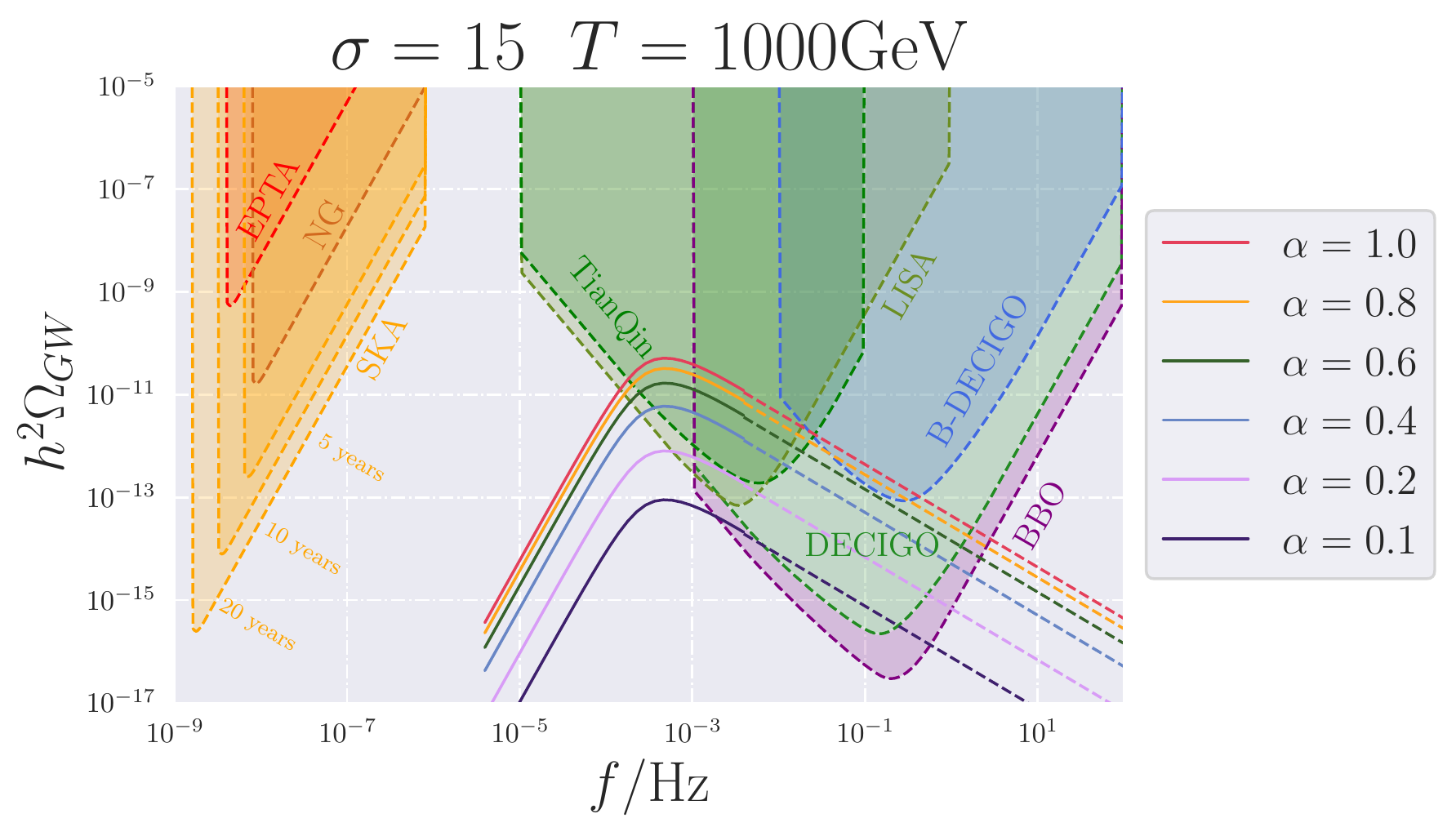}
\includegraphics[width=0.5\textwidth]{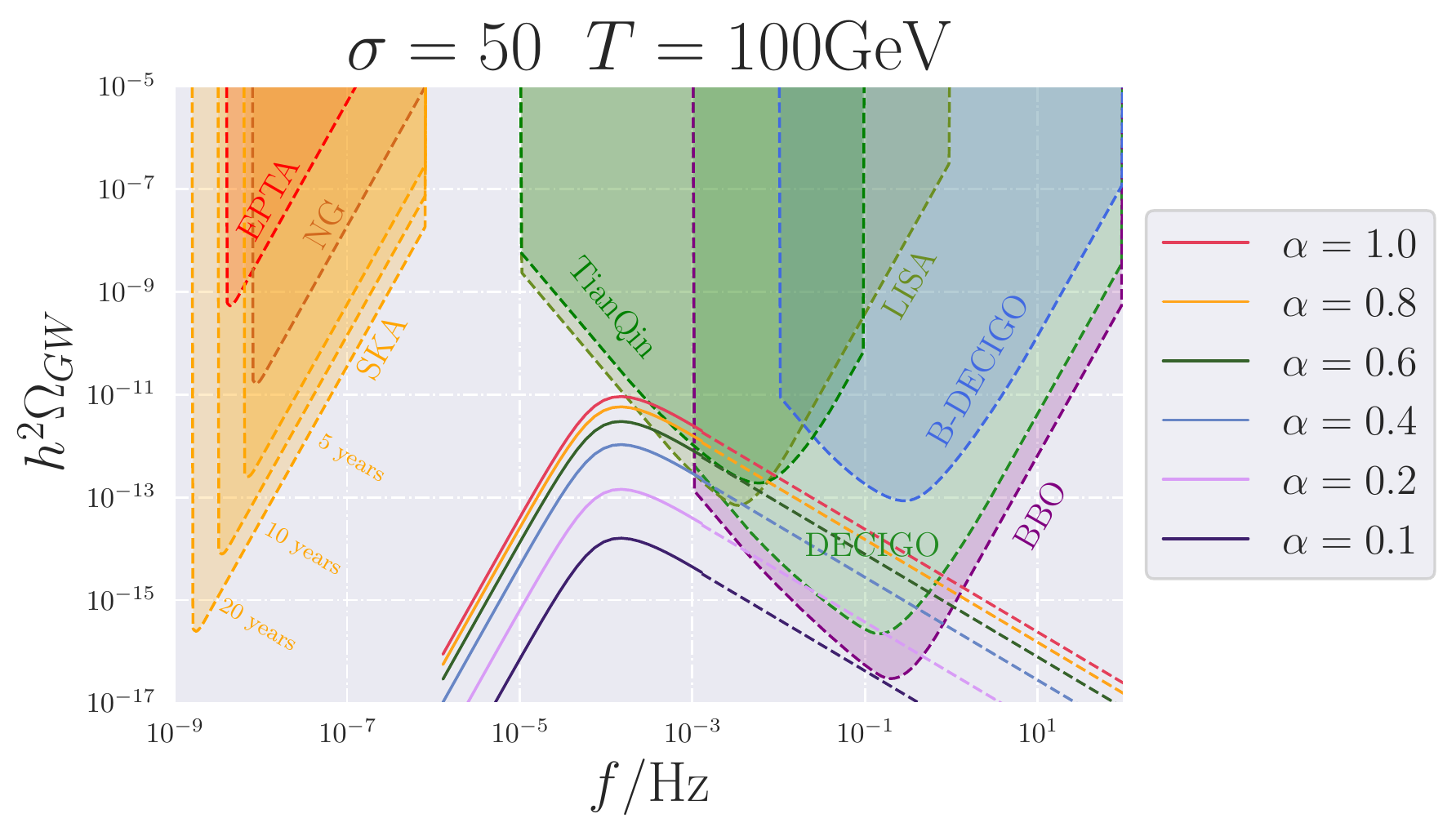}
\includegraphics[width=0.5\textwidth]{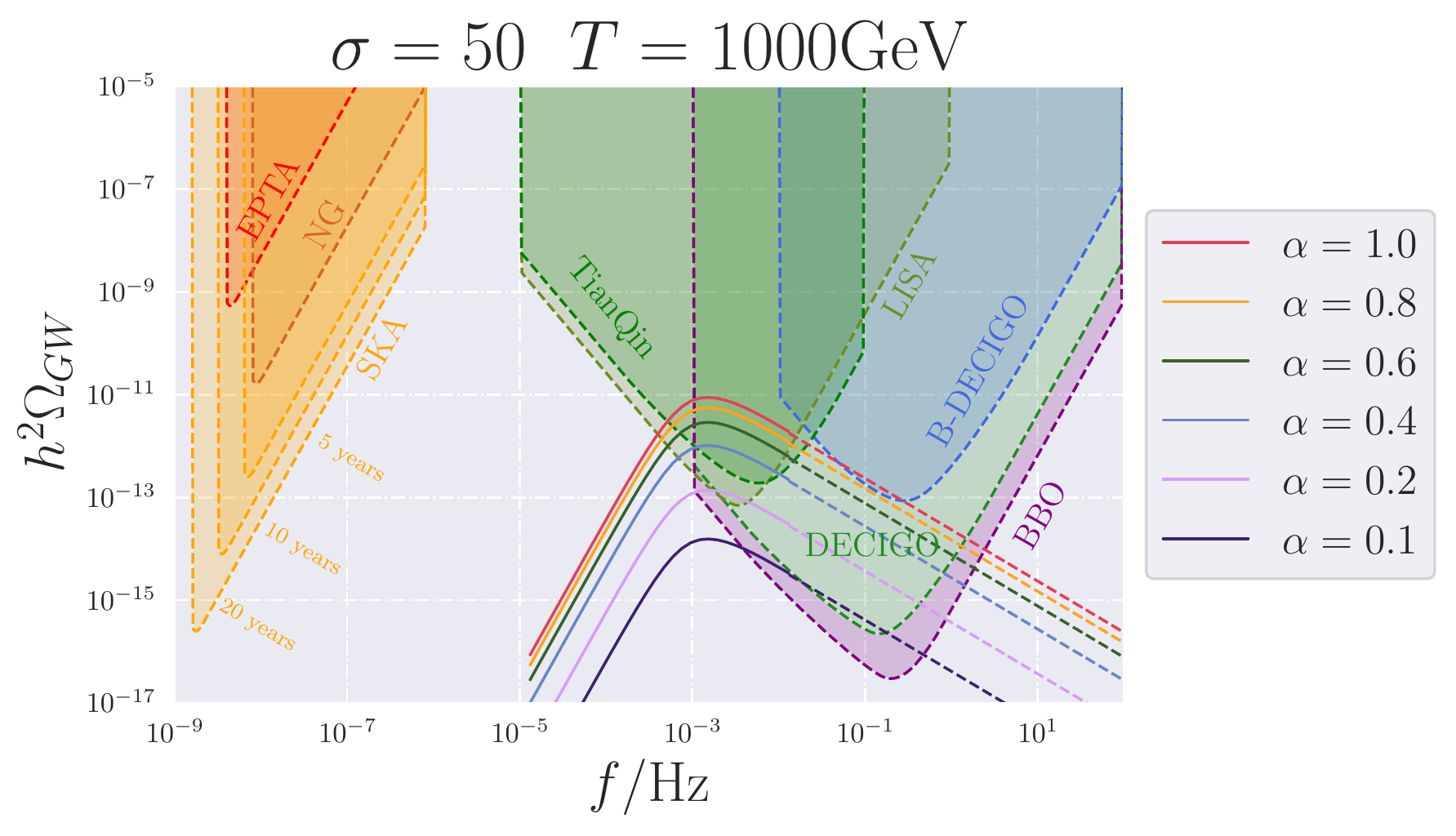}
\includegraphics[width=0.5\textwidth]{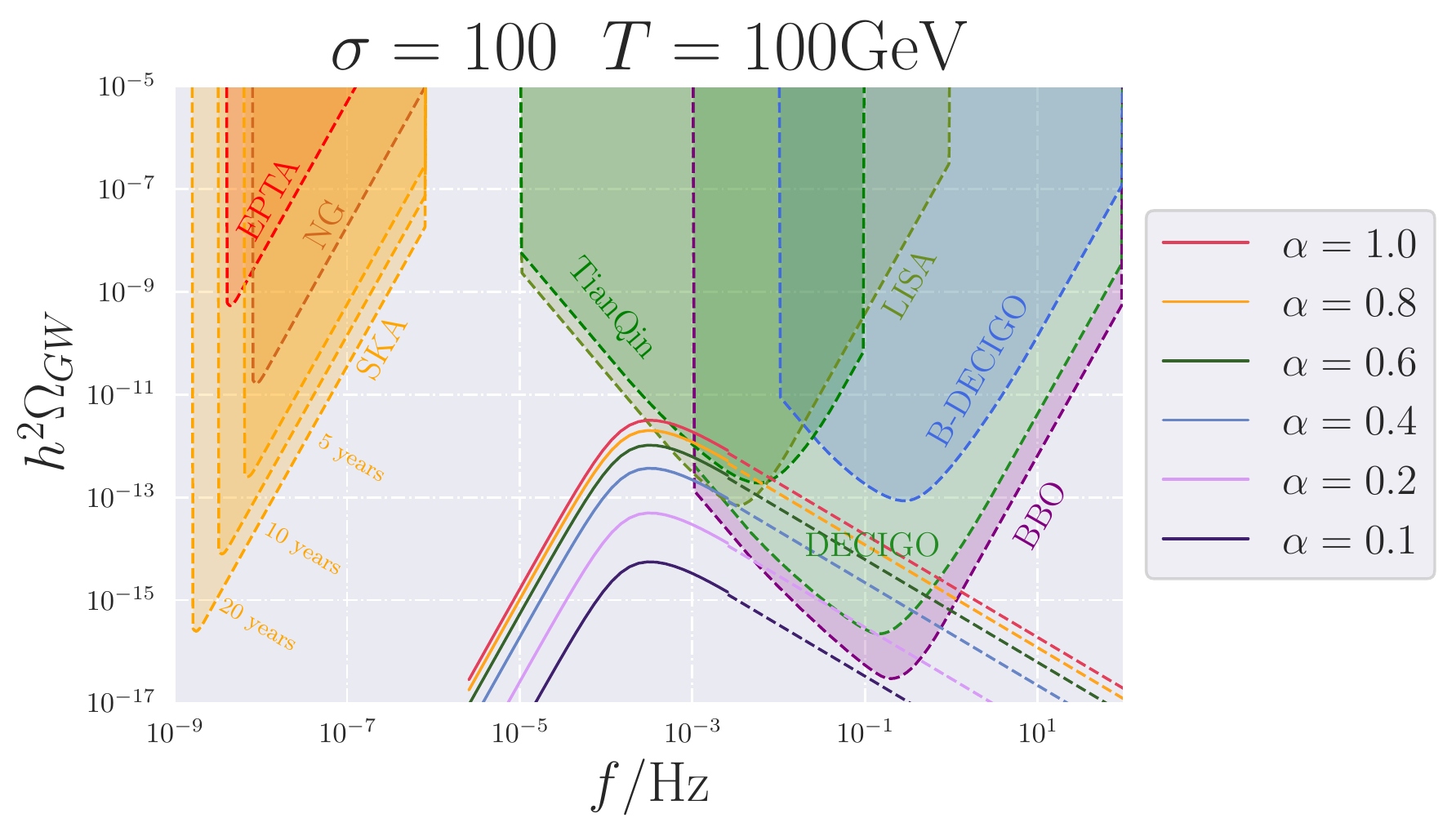}
\includegraphics[width=0.5\textwidth]{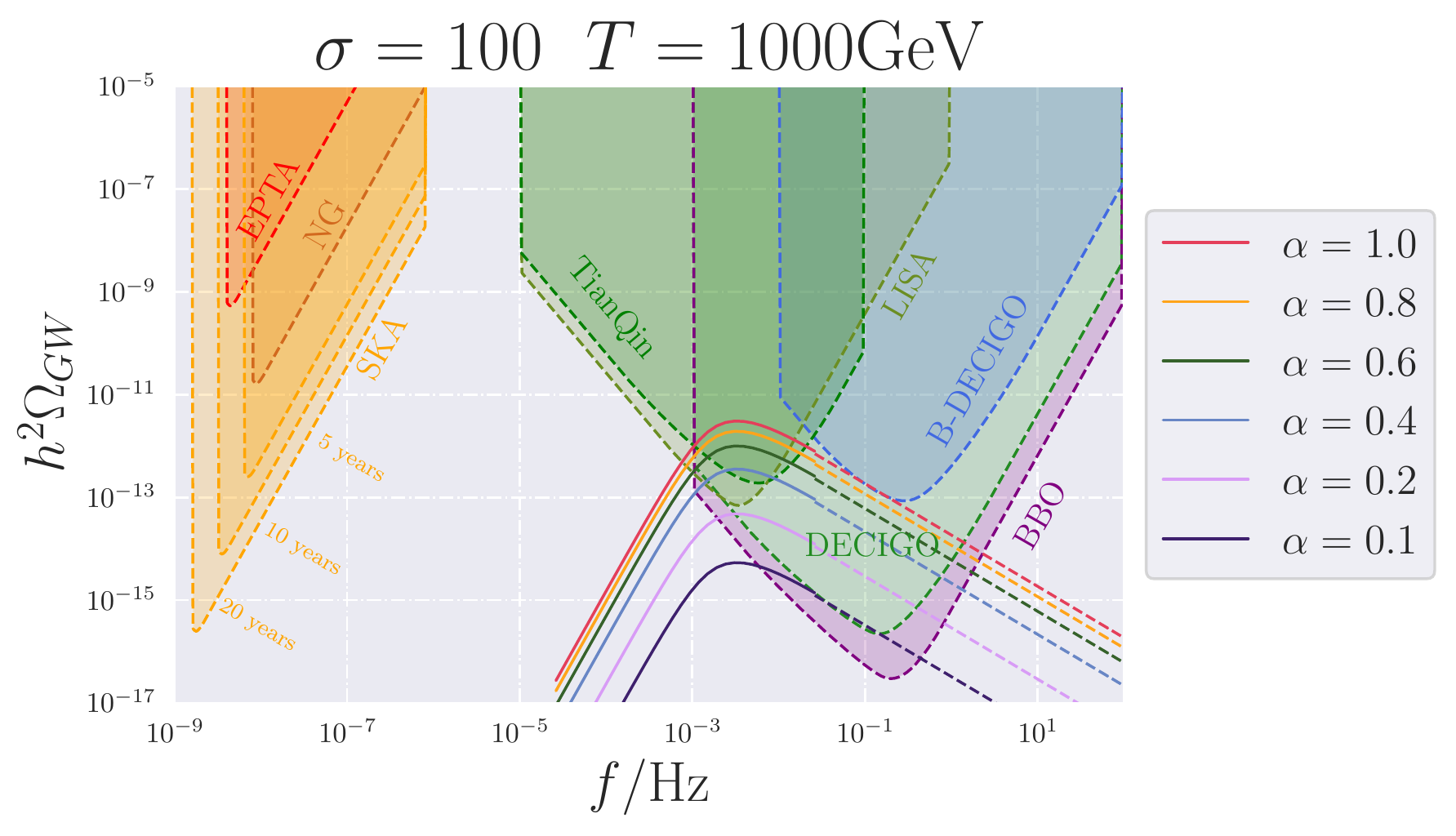}
\caption{Plot of GW spectra and PLI sensitivity curves\cite{PLI} of EPTA\cite{EPTA}, NanoGrav\cite{NG}, SKA\cite{SKA}, TianQin\cite{TianQin}, DECIGO\cite{DECIGO}, LISA\cite{LISA}, B-DECIGO\cite{DECIGO}, DECIGO\cite{DECIGO} and BBO\cite{BBO}. At here, we didn't choose any specific model of phase transitions, so $\sigma, T, \alpha$ are three free variables which can change without restrictions. The solid lines are calculated by numerically integrating and dotted lines are given by fitting.}
\label{fig:sensitivity}
\end{figure*}

Since GW spectrum is indeed depressed compared to Jinno and Takimoto's estimation in Minkowski spacetime when we consider a more realistic spacetime background and also take the finite duration of phase transitions into account, we are supposed to review the detectability of GW from bubble collisions in FLRW spacetime again by comparing GW spectra with PLI sensitivity curves of GW detectors. Out of the frequency band of GW we are considering about, we neglect the terrestrial-based detectors like LIGO and Virgo whose sensitive band is around hundred hertz. We choose five space-borne GW detector proposals at here including TianQin\cite{TianQin}, LISA\cite{LISA}, DECIGO\cite{DECIGO}, B-DECIGO\cite{DECIGO} and BBO\cite{BBO} and three Pulsar Timing Arrays including EPTA\cite{EPTA}, NanoGrav\cite{NG} and SKA\cite{SKA}. The results are shown by FIG.\ref{fig:sensitivity}.

According to FIG.\ref{fig:sensitivity}, we can find that although the maximum values of GW energy density do decrease a lot than in Minkowski spacetime, there still are many possible parameter combinations whose corresponding GW can be detected by TianQin, LISA, DECIGO and BBO which is really an impressing result. Even when $\sigma=100$, the energy density of GW from bubble collision is still large enough to be detected in the promising future. Again, we remind readers that for the little understanding of the precise physical processes happen in the era of phase transitions, we don't adopt any specific physical model to describe it. Under such a circumstance, $\sigma$, $\alpha$ and $T_\star$ are all free parameters. However, we still need to confirm that the phase transitions happen during the RD era, since our derivation is based on this elementary assumption.

\begin{figure*}[tbp]
\includegraphics[width=0.5\textwidth]{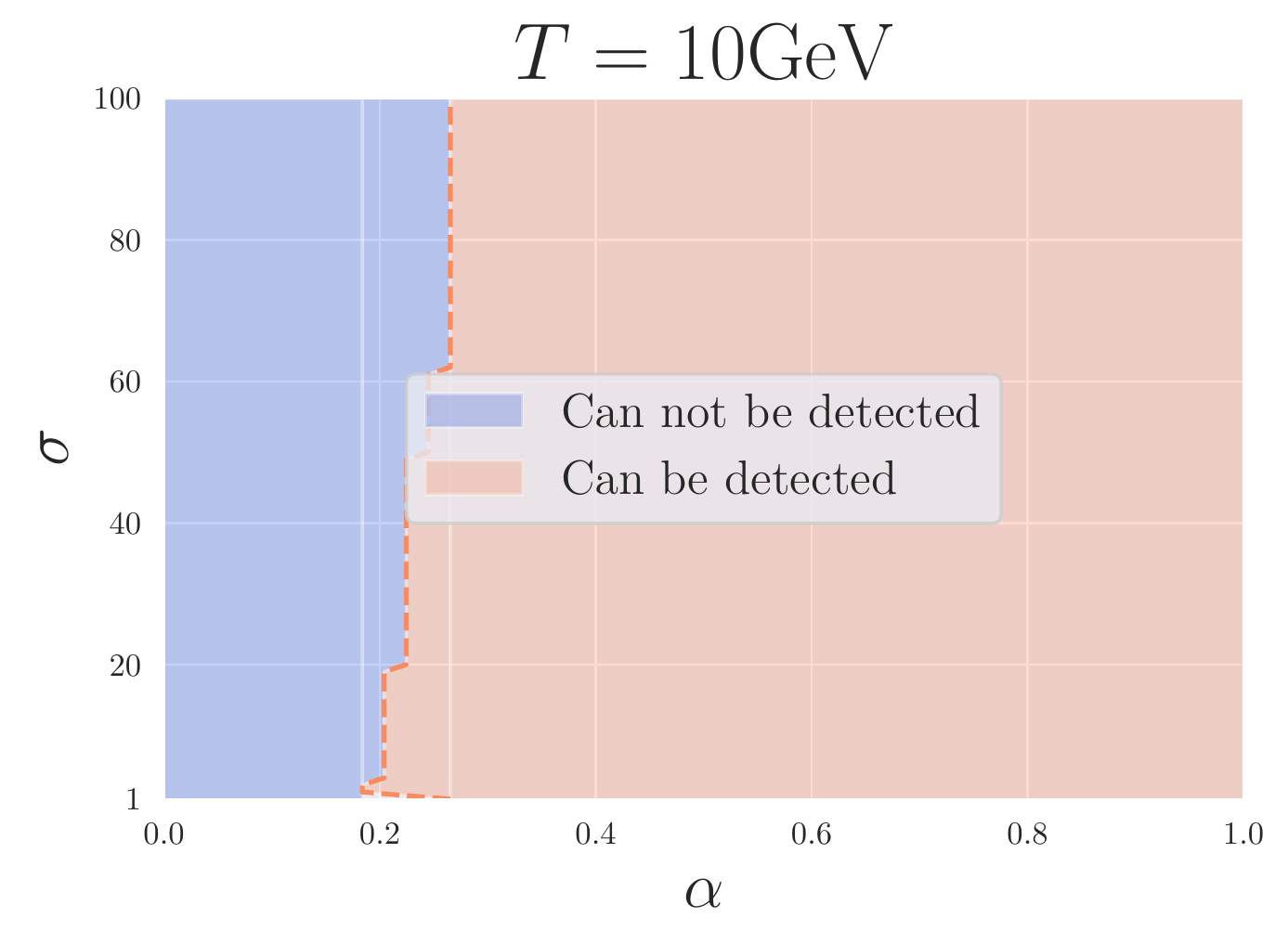}
\includegraphics[width=0.5\textwidth]{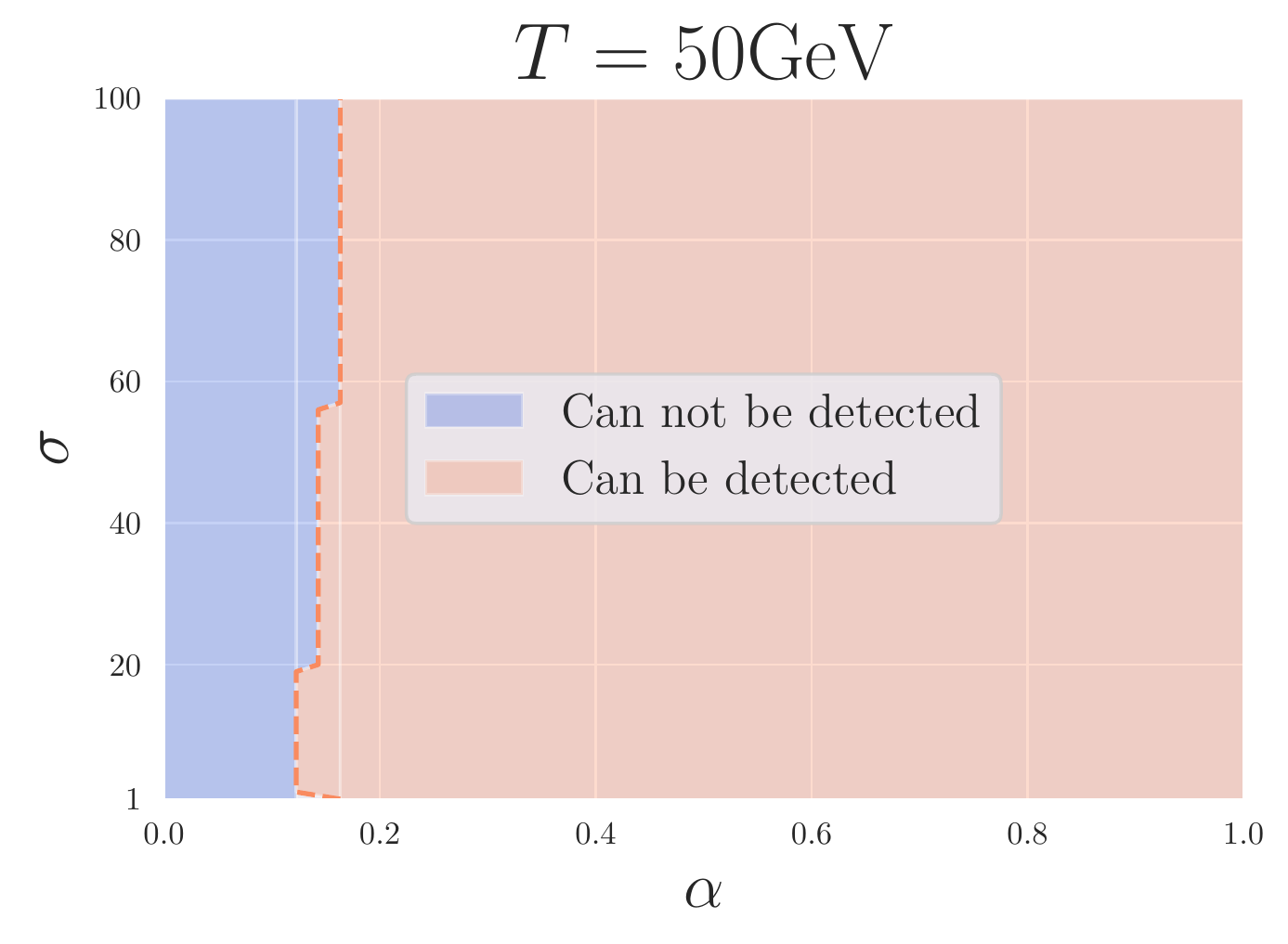}
\includegraphics[width=0.5\textwidth]{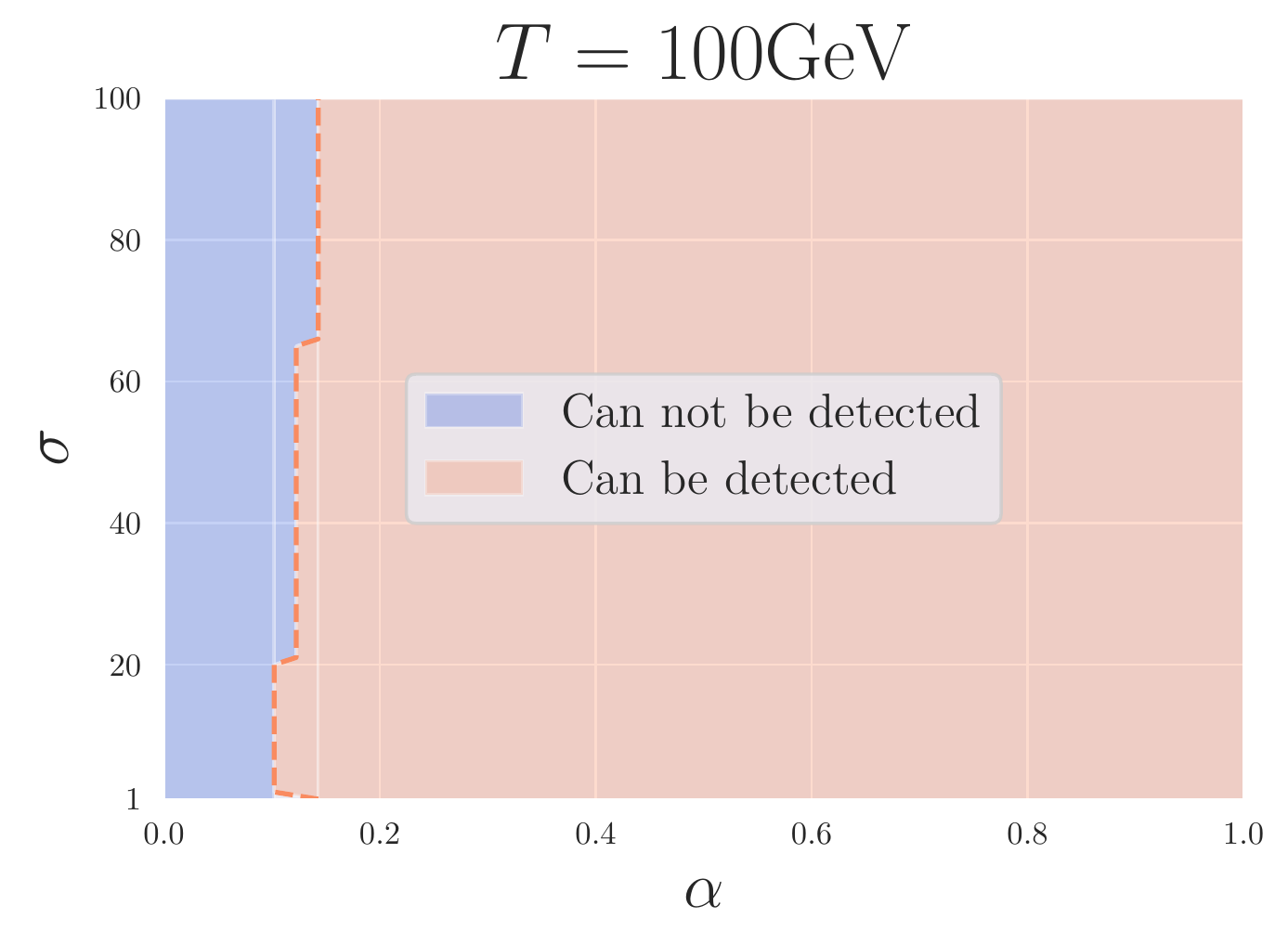}
\includegraphics[width=0.5\textwidth]{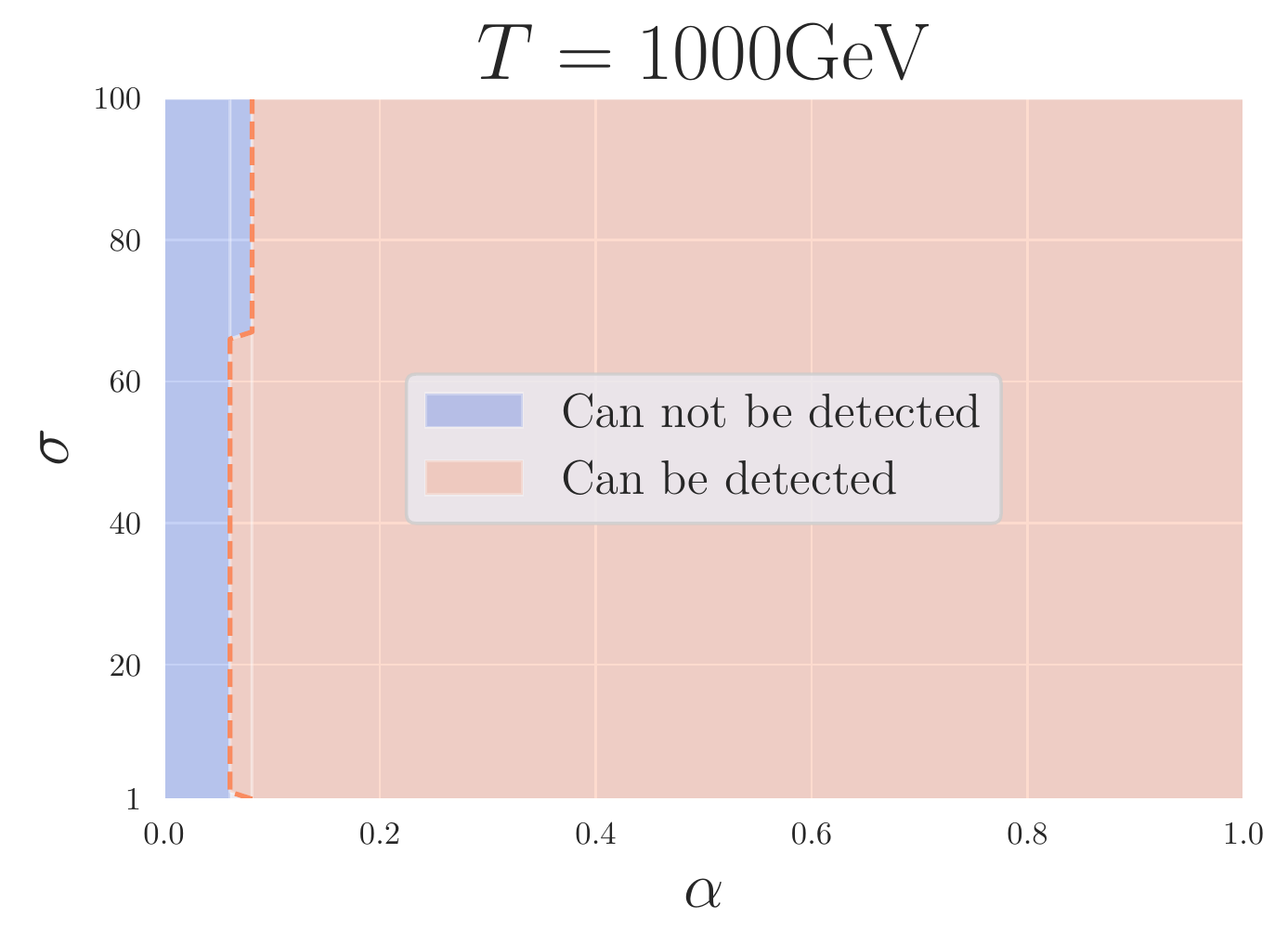}
\caption{Plotting of the parameter combinations in $\sigma-\alpha$ plane with different values of $T_\star$. The upper left Lyons blue area denotes the pairs of $(\sigma,\alpha,T_\star)$ whose corresponding GW can not be detected, while the lower right salmon pink area denotes the pairs of $(\sigma,\alpha,T_\star)$ whose corresponding GW can be detected.}
\label{fig:area}
\end{figure*}

Then we can go forward a small step to calculate the possible parameter combinations whose corresponding GW can be detected by the above detectors. Observing the shape of the sensitivity curves shown in FIG.\ref{fig:sensitivity}, we can find that as long as the curves of GW spectra fall above the sensitivity curves of BBO, this GW signal may be detected. Based on such a consideration, we can get the FIG.\ref{fig:area}.
Observing FIG.\ref{fig:area}, we can see that as the temperature after the phase transitions $T_\star$ becomes higher and higher, the area of parameter combinations whose corresponding GW can be detected (salmon pink parts in FIG.\ref{fig:area}) is becoming bigger and bigger, which is definitely what we could predict before the calculation. Recall Eq.(\ref{eq:Deltas}), Eq.(\ref{eq:Deltad}) and combine our numerical estimations, we can see in two situations where the values of $\Delta^{\text{F}}$ decrease monotonously. Firstly, $\Delta^{\text{F}}$ drops with the increase of $\sigma$, when $\sigma\gtrsim2$. Secondly, $\Delta^{\text{F}}$ drops with the decrease of $\sigma$, when $\sigma\lesssim2$. We can try to understand this phenomenon by considering about two limiting cases. Note that $\sigma$ denotes the speed of the phase transitions with respect to the speed of the expansion of the universe. If $\sigma\to\infty$ i.e. the phase transitions finish instantly, we can assume that every single spacetime point transform from the false vacuum state  into the true vacuum state at once where the spherical symmetry of the system is conserved and there are almost no uncollided bubbles left, so there isn't any detectable GW signal. On the other hand, if $\sigma\to0$ which means the phase transitions proceed very slowly, we still cannot expect a detectable GW signal for the low probability of bubble collisions. So we can expect that when $\sigma\simeq2$, the needed $\alpha$ is the smallest. For other cases, we all need a larger $\alpha$ to make sure the resulting GW could be detected. When $\sigma>2$, we can clearly see that the larger $\sigma$ is, the larger $\alpha$ is needed to have a detectable signal.

Now, let's discuss the difference between the four sub-figures of FIG.\ref{fig:area}. The only difference of the parameter for these four figures is the value of $T_\star$, we can discover that with the increase of $T_\star$, the corresponding salmon pink area becomes bigger. Actually, we can recall Eq.(\ref{eq:frequency}) to obtain a straightforward understanding. When $T_\star$ is larger than $100$ GeV, the frequency of GW will increase and move toward the sensitive band of space-borne GW detectors. Although the values of $h_0^2\Omega_{\text{GW}}$ will be lower with higher $T_\star$, the increase of the GW frequency can totally compensate for this negative impact.

To remind, at here we only adopt a special case of $\kappa(\alpha)$ for calculation whose results should be regarded as a benchmark rather than holding in all different cases. Nevertheless, the method we've taken is very straightforward and could be realized easily.
\section{\label{sec:6}Conclusion and Discussion}
In this work, we explored the energy spectra of GW from bubble collisions in FLRW spacetime during RD era with a finite duration of phase transitions. We adopted the analysis method proposed by Ryusuke Jinno and Masahiro Takimoto in 2017 and extended it to a conformal flat spacetime background. 

In the very first beginning, we gave a brief retrospection and summary of Jinno and Takimoto's work. For the convenience of readers, we also listed some useful equations and results of their work. Then by adopting their elegant approach, we derived the analytical expression of $\Delta^{\text{F}(s)}$ and $\Delta^{\text{F}(d)}$ as Eq.(\ref{eq:Deltas}) and Eq.(\ref{eq:Deltad}), respectively. Later, we discussed the behavior of $\Delta^{\text{F}}$ qualitatively and claimed that it would increase to the maximum value first and decrease later when $\tau$ increases. By the virtue of this property, we defined the so-called “effective duration” of phase transitions by finding the specific $\tau$ which could maximize $\Delta^{\text{F}}$. By numerical integration, we found that the maximum values of $\Delta$ depressed to only $10\%$ of Jinno and Takimoto's estimation in Minkowski spacetime when $\sigma\sim\mathcal{O}(10)$. Even when $\sigma$ is large as 100, the ratio $\max{(\Delta^{\text{F}})}/\max{(\Delta^{\text{M}})}$ is still as small as 64.5\%. The significant difference shows up out of two different reasons. Firstly, we chose a more realistic spacetime metric\textemdash FLRW metric\textemdash to describe an expanding universe. When the speed of phase transitions and the speed of the expansion of the universe was comparable, we could not neglect the “dilution” effect brought by the expansion of the universe. Secondly, we didn't assume the phase transitions start from $\eta_i=-\infty$ and end at $\eta_f=\infty$, on the contrary, we defined an effective duration of phase transitions to realize numerical estimation.

The big decrease in the energy density of GW might bring about some new challenges for future detection. So we thought it was necessary to review the detectablity of GW generated from bubble collisions in FLRW spacetime. We regarded $\sigma$, $\alpha$ and $T_\star$ as three free variables and compared the GW spectra corresponding to different parameter combinations with the PLI sensitivity curves of several gravitational waves detectors. After comparison, we found that albeit the overall decrease of GW spectra, there were still many possible parameter combinations whose corresponding GW spectra curves fell above the sensitivity curves of GW detectors. Considering the shape of the sensitivity curves of GW detectors, we could find that as long as the GW spectrum curves fell above the sensitivity curve of BBO, the corresponding GW might be detected by us in the future. Based on such a criterion, we got FIG.\ref{fig:area} and found that with the increase of $T_\star$, the possibility for us to detect GW increased as well. Note that not every point $(\sigma,\alpha,T_\star)$ in FIG.\ref{fig:area} is 
permitted by the theories of the first order cosmological phase transitions, provided we have chosen a specific model to describe the phase transitions, and then $\alpha, \sigma$ and $T_\star$ are all specified uniquely. However, since we have little knowledge about the real physics processes happened around transition time, we don't think it is necessary to choose a model. On the other hand, our result is very general. Given a specific model, we can calculate the energy spectrum and compare it with PLI sensitivity curves of GW detectors to testify the detectability of GW.

Besides these, there are many other works waiting to be done. Firstly, our derivation is limited to the phase transitions happen in RD era and the calculation in other eras are also demanded. Secondly, we only considered the situation where the bubble wall moved with the speed of light. Actually, in many models, the speed of the bubble wall doesn't need to be $c$, so it is a problem should be focused on the future research. 
Finally, we still used envelope approximation in our work, while there are several paper which have abandoned this approximation, see \textit{e.g.}\cite{Jinno_2019}. In a word, more detailed and elaborate discussions of GW from bubble collisions in FLRW spacetime still are needed to be done.

\section{Acknowledgements}
We thank Ryusuke Jinno in particular for his idea, reading our manuscript and fruitful discussion. This work is supported by the National SKA Program of China (2020SKA0120300), the National Key Research and Development Program of
China (No. 2020YFC2201400),and also the National Natural Science Foundation of China under Grants No. 11653002, No. 11875141 and the National Key R\&D Program of China (2021YFC2203100).
\nocite{*}
\bibliographystyle{plain}
\bibliography{main.bib}
\end{document}